\documentclass[lettersize,journal]{IEEEtran}

\usepackage{lettrine}
\usepackage{xcolor}
\usepackage{amsmath,amsfonts,amssymb}
\usepackage{algorithm}
\usepackage{array}
\usepackage[caption=false,font=normalsize,labelfont=sf,textfont=sf]{subfig}
\usepackage{textcomp}
\usepackage{stfloats}
\usepackage{url}
\usepackage{verbatim}
\usepackage{graphicx}
\usepackage{cite}
\usepackage{hyperref}

\definecolor{darkgreen}{RGB}{40,120,40}
\newcommand{\zhulatest}{\textcolor{darkgreen}}
\newcommand{\zhu}{\textcolor{black}}

\usepackage{algpseudocode}
\usepackage{svg} 
\usepackage{multirow}
\usepackage{booktabs}
\usepackage{graphicx} 
\usepackage{array}
\usepackage{siunitx}
\usepackage{placeins} 
\usepackage{capt-of}  
\usepackage{bbm}
\usepackage{hyperref}          
\usepackage{orcidlink}         

\begin{document}

\title{GARDEN: Efficient GPU-Accelerated Billion-Scale GNN Training via Caching-Sampling-Batching Co-Design}


\author{
    Qingrui Zhu\,\orcidlink{0009-0004-1138-4510}, Lei Liu\,\orcidlink{0009-0001-9665-8469}, Yushu Cai\,\orcidlink{0009-0009-1878-744X}, Kai Sheng\,\orcidlink{0009-0004-0234-4955}, Zhiwen Chen\,\orcidlink{0000-0003-1897-7840}, 
    ~\IEEEmembership{Member,~IEEE,}, Xin He\,\orcidlink{0009-0002-1481-3179}
    
    \thanks{Corresponding author: Xin He.}
    \thanks{Qingrui Zhu, Lei Liu, Yushu Cai, Kai Sheng, and Xin He are with the Guangzhou Institute of Technology, Xidian University, Guangzhou 510530, China (e-mail: zhuqingrui@stu.xidian.edu.cn; liulei2303@stu.xidian.edu.cn;  yushucai@stu.xidian.edu.cn; kaisheng@xidian.edu.cn; hexin@xidian.edu.cn).}%

    \thanks{Zhiwen Chen is with the School of Computer Science, Xiangtan University, Xiangtan 411105, China (e-mail: zhiwenchen@xtu.edu.cn).}%
}

\markboth{Journal of \LaTeX\ Class Files,~Vol.~14, No.~8, August~2021}%
{Shell \MakeLowercase{\textit{et al.}}: A Sample Article Using IEEEtran.cls for IEEE Journals}


\maketitle

\begin{abstract}
Graph Neural Networks (GNNs) have shown impressive performance in diverse tasks such as social network analysis and recommendation systems. However, scaling their training to billion-edge graphs is hindered by the memory wall: GPU memory capacity falls short of the massive storage requirements for node features and topology. Sampling-based GNN training frameworks mitigate this by streaming subgraphs from CPU to GPU, but they incur prohibitive PCIe data transfers and non-coalesced memory accesses, resulting in bandwidth underutilization and low compute efficiency. We propose GARDEN, a GPU-accelerated sampling-based system that co-designs caching, sampling, and batching to eliminate these bottlenecks. GARDEN introduces three key innovations: (1) an Adaptive High-Fidelity Cache Allocator (AHFCA) that combines offline reuse profiling and online hotness tracking to dynamically cache features and adjacency lists within a fixed GPU memory budget, avoiding static allocation waste; (2) a Grouping-based Parallel Sampler that reorders stochastic neighborhood accesses(GPS) to maximize warp occupancy and hide latency on power-law graphs; and (3) a Super-batch Degree Classifier(SDC) that fuses mini-batches into degree-homogeneous groups, enabling cache-friendly coalesced transfers and high GPU utilization. \zhulatest{Evaluations on real-world billion-scale graphs demonstrate that GARDEN achieves 4.54$\times$, 1.27$\times$, and 2.97$\times$ end-to-end training speedups over DGL-UVA, DUCATI, and FastGL, respectively, while maintaining model convergence and accuracy.}

\end{abstract}


\section{Introduction}

\lettrine[lines=2, lhang=0.1, loversize=0.1]{G}{raph} Neural Networks (GNNs)\cite{gnn} have emerged as a powerful tool for learning representations on graph-structured data, achieving state-of-the-art performance in diverse domains such as social network analysis\cite{social2}, recommendation systems\cite{HAGERec}, and molecular property prediction~\cite{molecular2}. However, scaling GNN training to real-world graphs with billions of edges (e.g., citation networks like \zhulatest{Papers100M (PA)}~\cite{2020OGB}, containing over 100 million nodes and 1.6 billion edges) poses a fundamental memory wall challenge. For instance, the Papers100M dataset requires hundreds of gigabytes for feature storage alone, far surpassing the limited capacity of GPU memory~\cite{2022GNNLab,2023NeutronOrch,2023DUCATI, Quiver}. This makes full-graph training~\cite{wholegraph} approaches impractical due to prohibitive memory footprints and CPU-GPU communication bottlenecks.

To bridge this gap, sampling-based GNN training has become a predominant paradigm, where subgraphs are sampled from the full graph and processed in mini-batches on GPUs~\cite{GraphSAGE,ladies}. In practice, sampling-based pipelines typically adopt a CPU-GPU heterogeneous setup: the CPU hosts the entire graph topology and node features, while the GPU handles forward and backward propagation. Each training iteration involves three stages: (1) sampling a subgraph on the host, (2) transferring the sampled topology and features to the GPU via PCIe (the memory I/O phase), and (3) executing GNN computations on the GPU. While this sampling-based approach enables  training on massive graphs, it introduces two critical system bottlenecks, particularly for billion-scale graph scenarios. First, frequent large-scale data transfers between host and device incur high I/O costs. Second, the sampling process itself involves irregular memory accesses and complex logic, leading to substantial latency on graphs with power-law degree distributions~\cite{DropNaE}. These bottlenecks are exacerbated in irregular topologies, where random accesses dominate and constrain overall training throughput.

Existing efforts  attempt to mitigate these overheads but fail to address them holistically. Early GPU-side caching systems like GNNLab~\cite{2022GNNLab} and FastGL~\cite{2024fastgl} rely on single-tier caching (storing only features), neglecting the structural data required for sampling. While recent dual-cache designs like DUCATI~\cite{2023DUCATI} co-cache features and topology, they suffer from static allocation inefficiencies and low-fidelity profiling. They allocate GPU memory based on offline heuristics that do not account for the diminishing returns of cache size or dynamic hotness shifts during training. Furthermore, these systems’ sampling kernels use generic thread scheduling that is ill-suited to the skewed degree distributions of real-world graphs, causing load imbalance and underutilized memory bandwidth.

To overcome these bottlenecks, we propose GARDEN, a GPU-accelerated sampling-based system that co-designs caching, sampling, and batching for efficient billion-scale GNN training. GARDEN introduces three key innovations: (1) an Adaptive High-Fidelity Cache Allocator (AHFCA) that integrates offline reuse profiling with online hotness tracking to dynamically manage GPU memory under fixed budgets, thereby mitigating inefficiencies from static allocation; (2) a Grouping-based Parallel Sampler that reorders stochastic neighborhood accesses to improve warp occupancy and hide memory latency on graphs with skewed degree distributions; and (3) a Super-batch Degree Classifier that fuses mini-batches into degree-homogeneous groups to enable cache-efficient data transfers and maximize GPU utilization. We implement GARDEN as an end-to-end framework and evaluate it on billion-scale real-world datasets including Papers100M and IGB-large. Our evaluation demonstrates that GARDEN effectively overcomes the memory wall and compute irregularity of billion-scale GNNs, enabling practical, high-throughput training of billion-edge GNNs on a single multi-GPU node.

In summary, this paper makes the following contributions:
\begin{itemize}
\item We propose the Adaptive High-Fidelity Cache Allocator, a dynamic memory management strategy that integrates end-to-end access statistics with runtime adaptation to maximize cache hits under strict memory budgets.
\item We design a Grouping-based Parallel Sampler that optimizes thread scheduling for stochastic sampling, significantly reducing warp divergence and increasing computational throughput on skewed graphs.
\item We develop a Super-batch Degree Classifier that reorganizes training batches to exploit data locality, enhancing both PCIe transfer efficiency and GPU core utilization.
\item We build GARDEN, an end-to-end framework integrating these techniques. Our extensive evaluation on billion-scale real-world datasets show that GARDEN achieves speedups of 4.54$\times$ over DGL-UVA~\cite{dgl-uva}, 1.27$\times$ over DUCATI~\cite{2023DUCATI}, and 2.97$\times$ over FastGL~\cite{2024fastgl}, respectively, while maintaining model accuracy.
\end{itemize}

\section{Background}
\subsection{Graph Neural Networks}

Graph Neural Networks (GNNs) are deep learning models that operate on graph-structured data, capable of end-to-end learning for tasks including node classification, link prediction, and graph-level regression. Formally, let $\mathcal{G} = (\mathcal{V}, \mathcal{E})$ be a graph with node set $\mathcal{V}$ ($|\mathcal{V}|$ nodes) and edge set $\mathcal{E} \subseteq \mathcal{V} \times \mathcal{V}$ ($|\mathcal{E}|$ edges). Each node $v \in \mathcal{V}$ is associated with an initial feature vector $\mathbf{x}_v \in \mathbb{R}^{d_0}$ that encodes domain information, where $d_0$ is the dimensionality of the input node features.

GNNs employ an iterative message-passing scheme across $K$ layers.
At each layer $k$, a node $u$ aggregates information from its neighborhood $\mathcal{N}(u) = \{ v \mid (v, u) \in \mathcal{E} \}$ using a differentiable message function $\phi^{(k)}(\cdot)$:
\footnotesize
\begin{equation}
    \mathbf{m}_u^{(k)} = \textsc{Aggregate}^{(k)} \left( \left\{ \phi^{(k)}\left( \mathbf{h}_v^{(k-1)}, \mathbf{h}_u^{(k-1)} \right) \mid v \in \mathcal{N}(u) \right\} \right)
\end{equation}
\normalsize
where $\mathbf{h}_u^{(k-1)}$ is the hidden state of $u$ from the previous layer, and $\textsc{Aggregate}$ is a permutation-invariant function such as sum, mean, or max.

The node then updates its representation by combining its previous state with the aggregated message:
\begin{equation}
    \mathbf{h}_u^{(k)} = \textsc{Update}^{(k)}\left(\mathbf{h}_u^{(k-1)}, \mathbf{m}_u^{(k)}\right)
\end{equation}
where $\textsc{Update}^{(k)}(\cdot)$ is typically a learnable function such as an MLP or GRU. After $K$ layers, each node’s embedding $\mathbf{h}_u^{(K)}$ captures both topological information and feature content from its $K$-hop neighborhood, enabling effective graph reasoning.





\subsection{Sampling-based GNN Training with Caching}
\zhulatest{Training GNNs on large-scale graphs is challenging because the number of involved neighbors grows rapidly with model depth, making full-neighbor aggregation memory-intensive and impractical for billion-edge graphs. Sampling-based training addresses this problem by selecting a fixed number of neighbors at each layer, thereby bounding the size of the computation graph and enabling memory-efficient mini-batch training.}

\begin{figure}
  \centering
  \includegraphics[width=\linewidth]{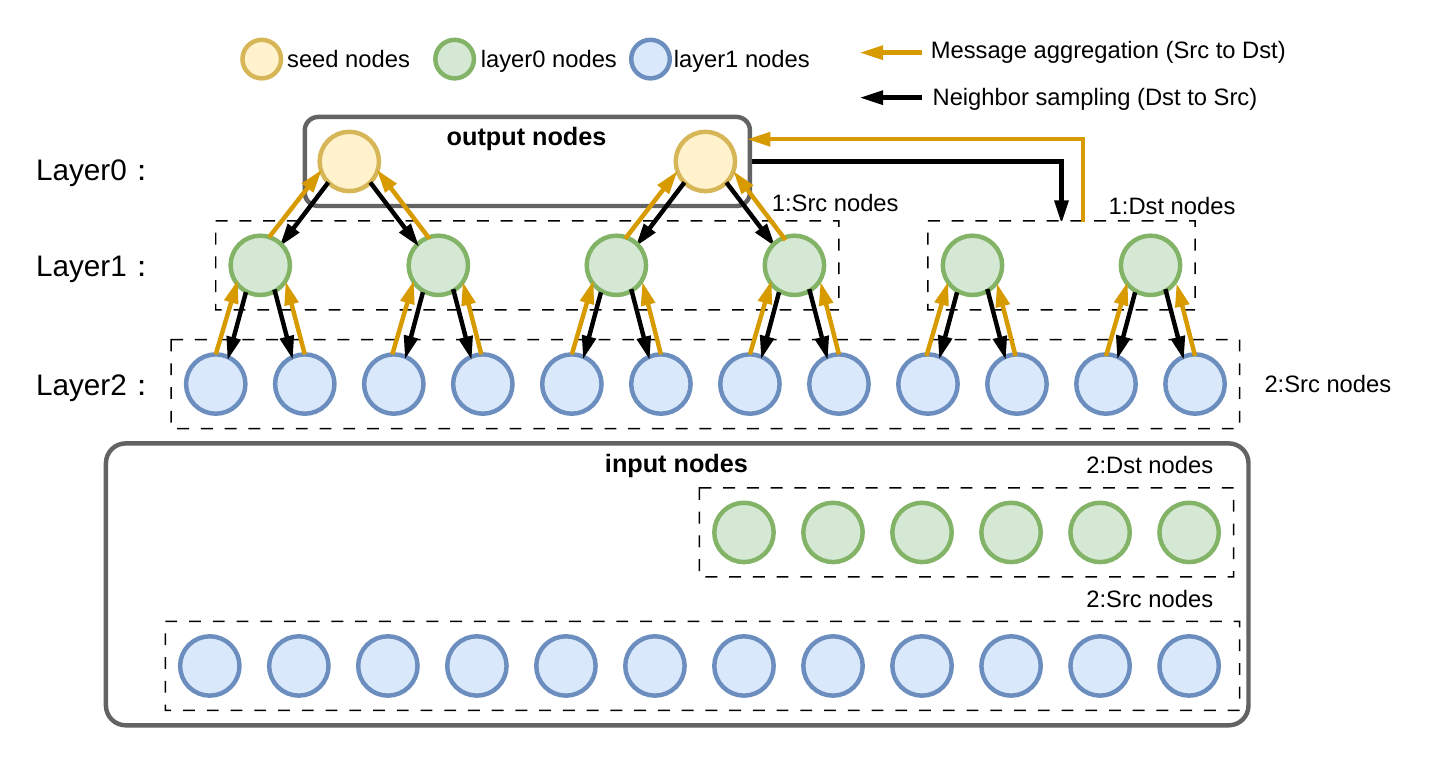}
  \caption{\zhulatest{Sampling-based GNN training framework with a per-layer fanout of [2,2]. Each layer-wise block connects source (src) nodes to destination (dst) nodes.}}
  \label{fig:sampling-background}
\end{figure}

\zhulatest{As illustrated in Figure~\ref{fig:sampling-background},
neighbor sampling begins with a set of seed nodes
$\mathcal{V}_0$ at Layer~0 and recursively expands toward the
outer layers according to the specified fanouts. At sampling
layer $\ell$, the node set $\mathcal{V}_{\ell-1}$ serves as the
Destination (Dst) set, whose representations are the aggregation
targets of that layer. The sampled neighbors are then added to
form the corresponding Source (Src) set,
$\mathcal{V}_{\ell}
=\mathcal{V}_{\ell-1}
\cup\mathcal{N}_{\ell}\!\left(\mathcal{V}_{\ell-1}\right)$.
Therefore, Src and Dst denote layer-specific roles rather than
disjoint node categories, and the Dst set of each layer is
contained in its corresponding Src set.}

\zhulatest{The two endpoint sets play different roles in
sampling-based training. The seed set $\mathcal{V}_0$ is the
\emph{output-node set}, because the model produces final
representations and evaluates predictions and losses for these
nodes. In contrast, $\mathcal{V}_k$, obtained after $k$ sampling
layers, is the \emph{input-node set}, because the raw features of
all nodes in this complete Src set must be fetched before GNN
computation. Consequently, neighbor sampling expands from the
output nodes at Layer~0 to the input nodes at Layer~$k$, whereas
message aggregation proceeds in the reverse direction, from the
input nodes through the intermediate layers to the output nodes.}

From a system's viewpoint, recursive sampling converts irregular propagation into sparse, localized operations, facilitating asynchronous fetching and dynamic scheduling. However, implementations vary in efficiency. PyTorch Geometric (PyG)\cite{pyg} relies on CPU sampling, limiting scalability for massive graphs. DGL accelerates sampling on GPUs but incurs overhead from synchronization and host-device transfers. Many frameworks also suffer from low cache hit rates due to graph irregularity, exacerbating bottlenecks. \zhulatest{Single-feature-cache systems such as PaGraph~\cite{PaGraph}, Data Tiering~\cite{DataTiering}, AliGraph~\cite{AliGraph2019}, and GNNLab~\cite{2022GNNLab} cache node features to reduce data transfers but overlook adjacency structures, constraining performance for large subgraphs.} FastGL employs single-cache optimization and reduces synchronization but introduces online overhead and fails to handle random accesses effectively. Dual-cache systems like DUCATI and DCI\cite{2025DCI} improve on this by separately caching features and adjacency, better suiting divergent access patterns. Yet, their static allocation and coarse profiling limit adaptability to dynamic workloads, often leading to inefficiencies in billion-scale scenarios. GNNBoost~\cite{GNNBOOST} introduces a cost-model-driven cache allocation framework for jointly optimizing topology and feature caching, but it suffers from several limitations. Its additional RNGS cache competes for limited GPU memory and its sampling strategy fails to hide random access latency, leaving significant I/O bottlenecks. Moreover, its reliance on offline models limits adaptability to dynamic workload changes. This highlights the need for more adaptive caching and sampling mechanisms.
\begin{figure}[h]
  \centering
  \includegraphics[width=\linewidth]{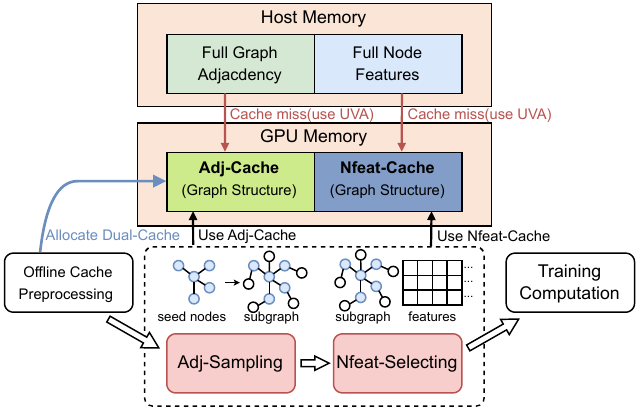}
  \caption{Architecture of the dual-cache GNN training system, coordinating feature and adjacency caches for efficient sampling.}
  \label{fig:dualcache-background}
\end{figure}

Next, we outline the Dual-Cache System (DCS) pipeline, as shown in Figure~\ref{fig:dualcache-background}. DCS comprises three stages. In offline preprocessing, it profiles access frequencies for adjacency lists and features, computes utility as frequency divided by size, and greedily populates Adj-Cache and Nfeat-Cache under GPU memory constraints. During training, Adj-Sampling queries Adj-Cache for lists, falling back to Unified Virtual Addressing (UVA)\cite{UVA}for host misses in Compressed Sparse Column (CSC) format, enabling low-latency sampling despite irregularity. Nfeat-Selecting follows suit for features, using pinned host memory on misses, with dense storage for bandwidth efficiency. DCS decouples caching for Adj-Sampling (latency-bound due to sparse pointer-chasing) and Nfeat-Selecting (bandwidth-bound from contiguous reads), avoiding single-cache trade-offs and improving throughput and utilization.

\subsection{Capability-Oriented Comparison}

\zhulatest{Table~\ref{tab:cache-capability} summarizes explicit design capabilities rather than measured performance. The comparison covers offline profiling, Dual-Cache organization, recent residual adaptation, sampling co-design, and batching co-design. A $\checkmark$ indicates that the corresponding capability is explicitly incorporated into the system design, whereas a $\times$ indicates that it is not included in the stated design scope.}

\zhulatest{GARDEN's main design distinction is the integration of these capabilities in one end-to-end framework. Full-link offline profiling and Dual-Cache organization preserve stable global reuse, while the bounded Dynamic-Cache compensates for recently recurring residual hot nodes missed by the static cache. GARDEN further couples cache management with sampling and batching through GPS and SDC, connecting cache decisions with execution organization. Among the listed systems, GARDEN is the only framework covering all five design dimensions. Table~\ref{tab:cache-capability} summarizes this capability coverage and clarifies the design scope rather than providing a direct performance ranking.}

\begin{table}[H]
  \centering
  \scriptsize
  \setlength{\tabcolsep}{0.5pt}
  \caption{\zhulatest{Capability-oriented comparison of cache organization and system co-design.}}
  \label{tab:cache-capability}
  {\color{darkgreen}
  \begin{tabular}{>{\raggedright\arraybackslash}p{0.22\columnwidth}*{5}{>{\centering\arraybackslash}p{0.13\columnwidth}}}
    \toprule
    Work & \shortstack{Offline\\profile} & \shortstack{Dual-\\Cache} & \shortstack{Recent\\residual\\adapt.} & \shortstack{Sampling\\co-design} & \shortstack{Batching\\co-design} \\
    \midrule
    GNNLab~\cite{2022GNNLab} & $\checkmark$ & $\times$ & $\times$ & $\times$ & $\times$ \\
    DUCATI~\cite{2023DUCATI} & $\checkmark$ & $\checkmark$ & $\times$ & $\checkmark$ & $\times$ \\
    FastGL~\cite{2024fastgl} & $\checkmark$ & $\times$ & $\times$ & $\checkmark$ & $\times$ \\
    GNNBoost~\cite{GNNBOOST} & $\checkmark$ & $\checkmark$ & $\times$ & $\checkmark$ & $\times$ \\
    GARDEN & $\checkmark$ & $\checkmark$ & $\checkmark$ & $\checkmark$ & $\checkmark$ \\
    \bottomrule
  \end{tabular}
  }
\end{table}
\FloatBarrier




\section{Motivation}
To understand the fundamental bottlenecks in billion-scale GNN training, we conduct a series of micro-benchmarks on state-of-the-art frameworks (DGL~\cite{2019DGL}, DUCATI~\cite{2023DUCATI}, and FastGL~\cite{2024fastgl}) using massive datasets such as Papers100M~\cite{2020OGB}, IGB-large~\cite{igb} and UK-2006-05~\cite{uk}. Our analysis reveals two primary systemic inefficiencies suffered by existing approaches.

\subsection{The Static Cache Dilemma: Mismatched Allocation and Diminishing Returns}
GNN training on billion-scale graphs demands efficient data management to handle massive node features and irregular graph structures. Existing caching strategies, however, fall short in adapting to diverse sampling algorithms and dynamic workloads. Single-cache systems, such as GNNLab and FastGL, typically cache only node features, neglecting graph topology like adjacency lists. This leads to high overhead in the sampling phase, where repeated CPU-GPU transfers dominate runtime.These single-cache approaches also underutilize GPU memory, often leaving several gigabytes of VRAM idle due to their static handling of GNN workloads. Dual-caching mechanisms, like DUCATI, co-cache node features and adjacency lists to improve data reuse and reduce I/O. Yet, even these systems struggle with billion-scale graphs, as static caching of node features mismatches on-demand training needs.

\begin{figure}
  \centering
  \includegraphics[width=\linewidth]{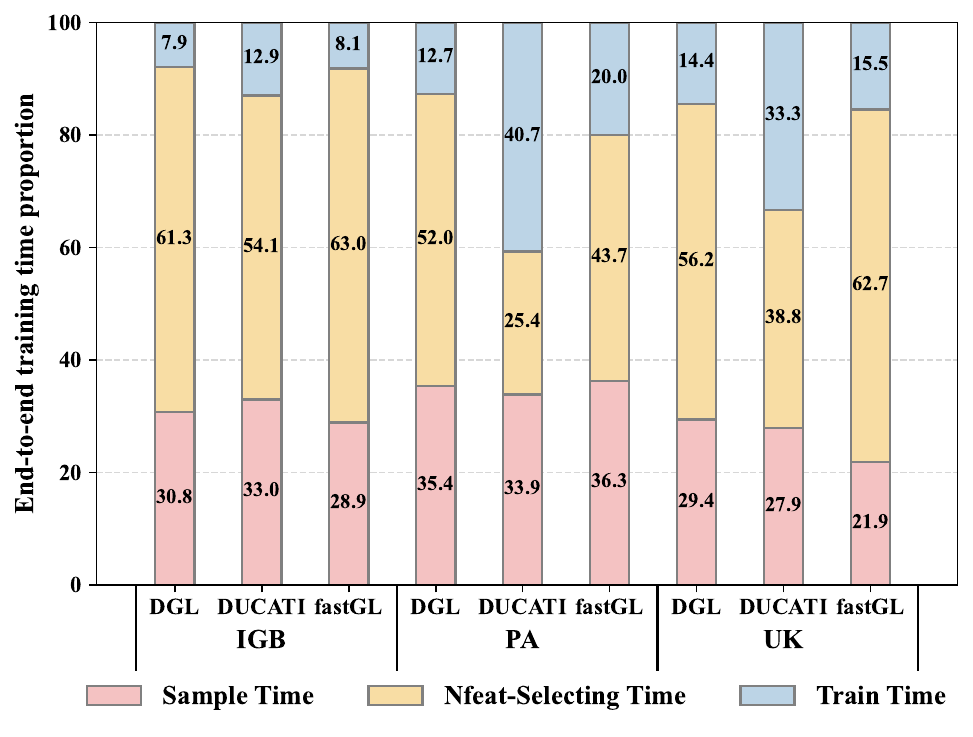}
  \caption{Execution time breakdown for GraphSAGE training on billion-scale graph dataset using NVIDIA RTX 3090 GPU.}
  \label{fig:breakdown}
\end{figure}


\begin{table}[htbp]
\centering
\caption{Feature Cache Hit Rate (\%)}
\begin{tabular}{lccc}
\toprule
Dataset & Framework & Hit Rate (\%) \\
\midrule
\multirow{5}{*}{IGB-large} & {\color{darkgreen}GNNLab} & 49.42 \\
 & DUCATI & 61.34 \\
 & FastGL & 16.96 \\
 & {\color{darkgreen}PaGraph} & {\color{darkgreen}12.49} \\
 & {\color{darkgreen}AliGraph} & {\color{darkgreen}13.98} \\
\midrule
\multirow{5}{*}{Papers100M} & {\color{darkgreen}GNNLab} & 83.35 \\
 & DUCATI & 94.94 \\
 & FastGL & 73.94 \\
 & {\color{darkgreen}PaGraph} & {\color{darkgreen}64.92} \\
 & {\color{darkgreen}AliGraph} & {\color{darkgreen}70.34} \\
\midrule
\multirow{3}{*}{UK} & {\color{darkgreen}GNNLab} & 41.06 \\
 & DUCATI & 72.77 \\
 & FastGL & 8.72 \\
\bottomrule
\end{tabular}
\label{tab:hitrate}
\end{table}

As shown in Figure~\ref{fig:breakdown}, the Nfeat-selecting phase emerges as a dominant bottleneck in systems like FastGL and DUCATI across benchmarks such as Papers100M, IGB-large, and UK, consuming a substantial fraction (e.g., on average 50\%) of total runtime. This overhead arises from a fundamental mismatch between offline-preprocessed caches and runtime feature demands: offline frequency-based profiling fails to capture the stochastic nature of multi-hop sampling, \zhu{causing 'hot' features to be evicted or missed entirely during training, resulting in cache hit rates that are predominantly below 90\% across all evaluated billion-scale workloads}, as detailed in ~\autoref{tab:hitrate}. Enhancing cache allocation during preprocessing is essential to improve hit rates and utilization.

\begin{figure}[h]
  \centering
  \includegraphics[width=\linewidth]{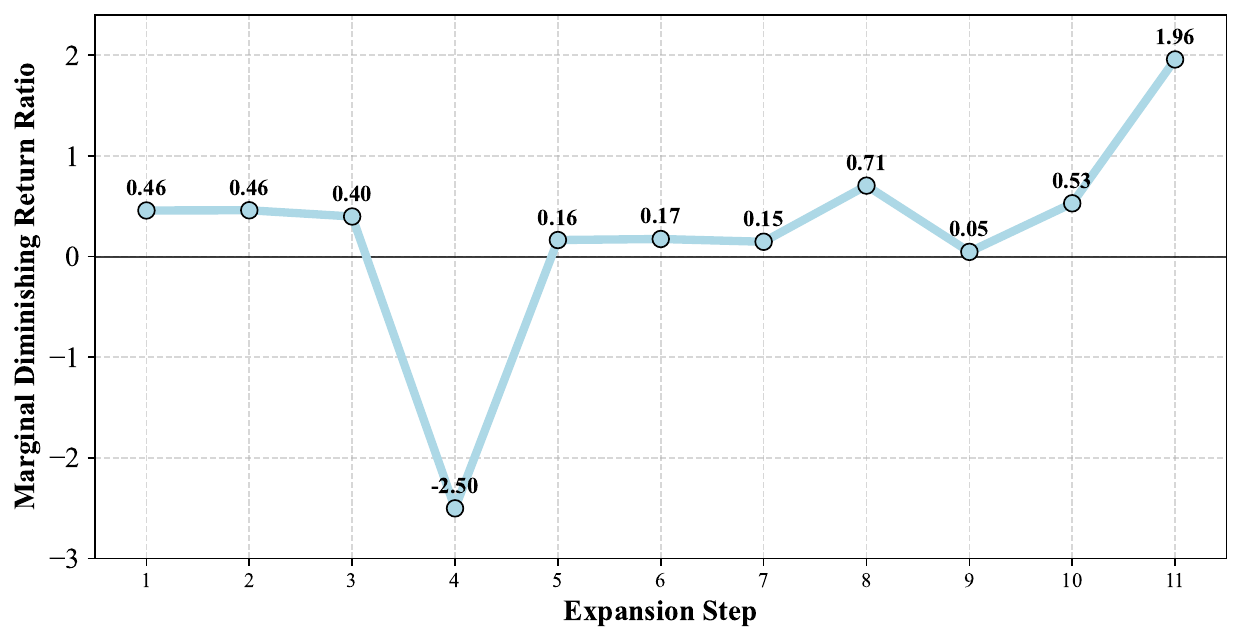}
  \caption{\zhulatest{Marginal diminishing-return ratio $r_i$ of DUCATI on the UK dataset over 11 cache expansion steps. Step $i$ corresponds to the transition from $(i+3)$ to $(i+4)$~GB; the $3\to4$~GB transition provides the reference marginal gain.}}
  \label{fig:cacheratio}
\end{figure}

GPU memory allocation further exacerbates inefficiencies. While software-managed feature caches reduce I/O latency, empirical analysis shows diminishing returns with increasing cache size. To quantify this, we define the marginal decay ratio as:
\begin{equation}
    r_i = 1 - \frac{\Delta_i}{\Delta_{i-1}}, \quad i = 1, 2, \dots, 11
\end{equation}


\zhulatest{Here, $\Delta_0$ denotes the reference latency reduction from 3 to 4~GB, and $\Delta_i$ denotes the latency reduction from $(i+3)$ to $(i+4)$~GB for $i=1,\dots,11$. Thus, Steps 1--11 correspond exactly to the 4$\to$5~GB through 14$\to$15~GB expansions shown in Figure~\ref{fig:cacheratio} and listed in Table~\ref{tab:cacheratio2}. A negative $r_i$ means that the current expansion stage provides a larger marginal benefit than the preceding stage. The 11-step DUCATI measurement exhibits non-monotonic marginal returns, including a pronounced fluctuation at Step 4. More importantly, all ratios from Steps 6--11 are positive, indicating that after Step 5 each further cache expansion provides a smaller marginal latency reduction than the preceding expansion. The large positive ratio of 1.96 at Step 11 further shows that DUCATI remains constrained by diminishing marginal returns as cache capacity increases.}


\textbf{Implication.} This insight reveals that static memory allocation to feature caches is suboptimal under fixed GPU budgets. Dynamic-Cache reallocation toward regions with the highest marginal gains could maximize training throughput. In essence, billion-scale GNN training requires optimizations that address static precaching limitations and enable adaptive memory use.

\subsection{The Sampling Bottleneck: Uncoalesced Access and Bandwidth Underutilization}
\begin{figure}[h]
  \centering
  \includegraphics[width=\linewidth]{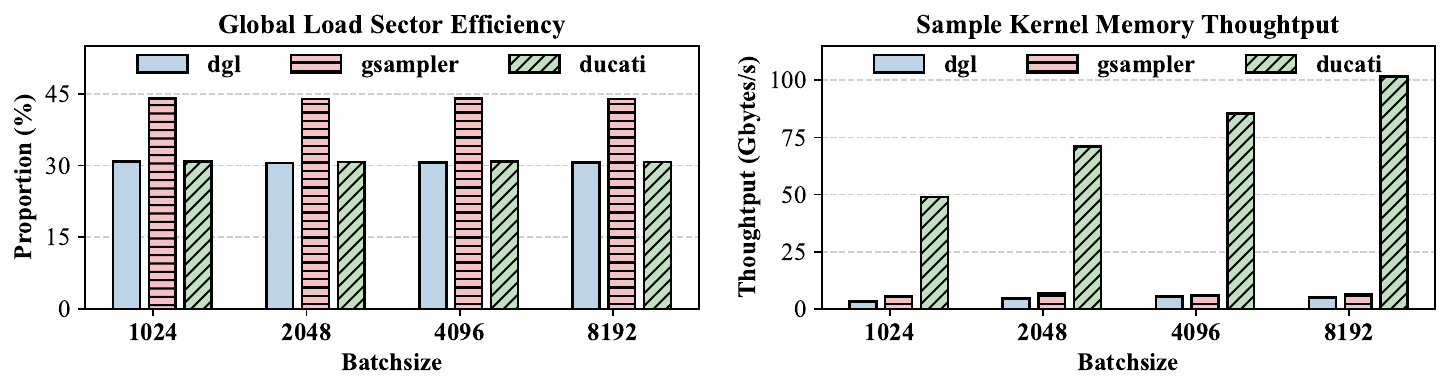}
  \caption{Fine-grained performance metrics measured using NVIDIA Nsight Compute on the Papers100M dataset.}
  \label{fig:global_load}
\end{figure}

Sampling in large-scale GNN training introduces severe bottlenecks due to the irregular topology of billion-scale graphs. Irregular access patterns hinder GPU parallelism, while node ID remapping and inter-thread dependencies impose synchronization overhead, reducing throughput. Profiling sampling kernels in frameworks like DGL, GSampler, DUCATI, and FastGL using Nsight Compute\cite{nsightcompute} reveals persistent inefficiencies. A key metric, Global Load Sector (GLS) Efficiency, measures bandwidth utilization by the ratio of valid data bytes to total sector size (e.g., 128 bytes on Ampere architectures). Values near 100\% indicate coalesced accesses, while low values (e.g., below 25\%) signal wasted bandwidth from uncoalesced patterns.

As illustrated in \autoref{fig:global_load}, even DUCATI, which employs adjacency list caching, shows poor GLS efficiency despite higher memory throughput, indicating that most transferred data remains unused.  All frameworks achieve throughput far below hardware limits (e.g., 1008 GB/s on RTX 3090 Ti), confirming sampling as a critical system bottleneck. We identify three root causes for these inefficiencies. First, the inherent randomness of graph sampling generates uncoalesced memory accesses that bypass GPU cache prefetching mechanisms. Second, existing caching strategies optimize data transfer but fail to address fundamental locality issues. Third, thread divergence and load imbalance in highly parallel GPU architectures exacerbate these problems, particularly for power-law graphs where degree distributions vary by orders of magnitude. 

\textbf{Implication.} These bottlenecks collectively constrain GNN training efficiency, necessitating a co-designed approach that optimizes caching, sampling, and batching simultaneously. Our proposed GARDEN framework addresses these limitations through three synergistic innovations: an Adaptive High-Fidelity Cache Allocator for dynamic memory management, a Grouping-based Parallel Sampler for optimized GPU utilization, and a Super-batch Degree Classifier for efficient batch processing.
\begin{figure}[h]
  \centering
  \includegraphics[width=\linewidth]{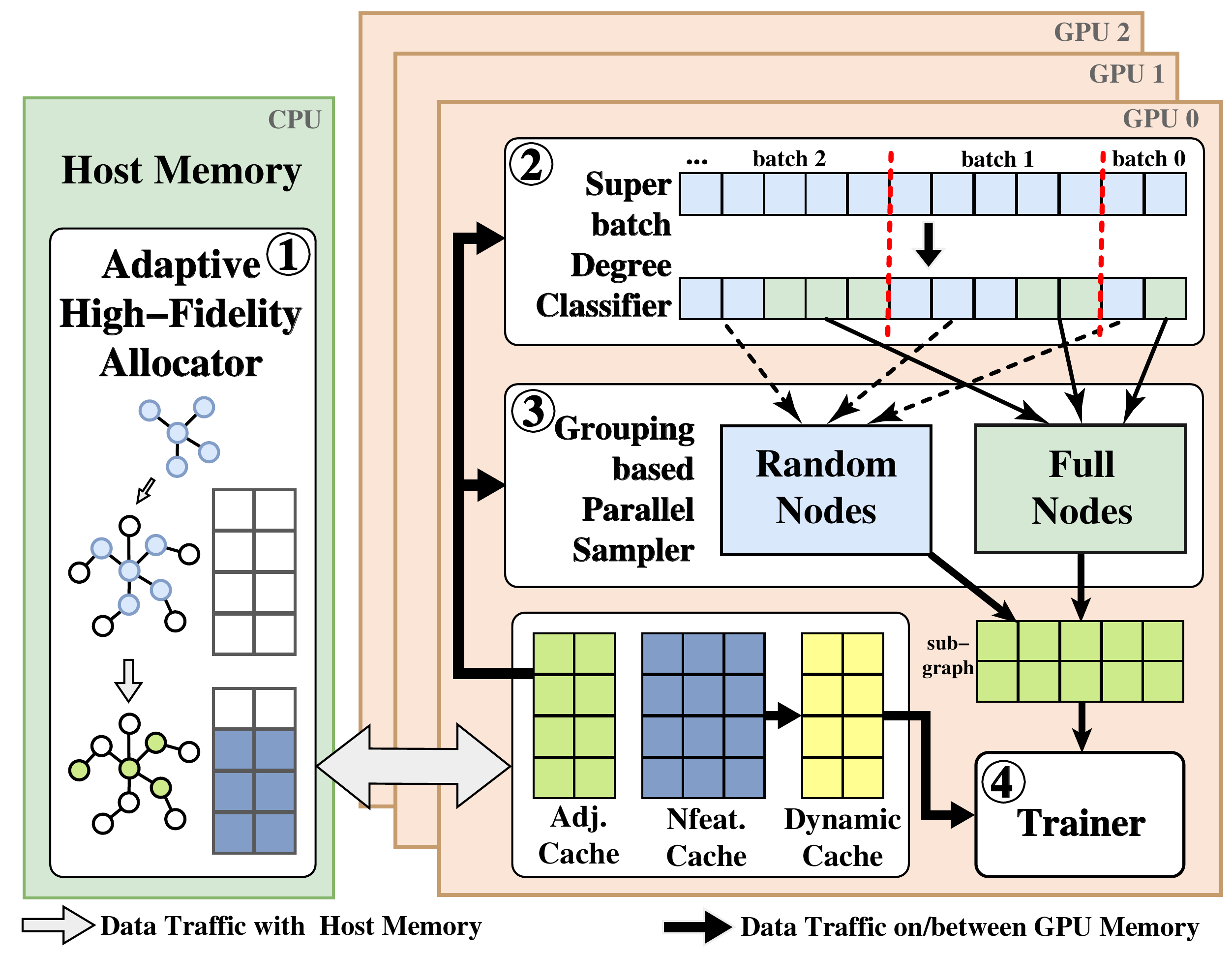}
  \caption{Overall architecture of GARDEN}
  \label{fig:overview}
\end{figure}

\section{Design}
\subsection{Overview}
We present GARDEN, an efficient sampling-based system for billion-scale GNN training. The system holistically co-optimizes caching, sampling, and batching to maximize training throughput on multi-GPU platforms. Figure~\ref{fig:overview} illustrates GARDEN's architecture. On the host side, the Multiple Dual Cache Allocator builds a dual-cache structure in each GPU's global memory, consisting of an adjacency cache (Adj-Cache) and a feature cache (Feat-Cache). Using offline pre-sampling statistics on node access frequencies, the allocator reassigns compact IDs to high-frequency nodes and populates the caches with their adjacency lists and feature vectors. This minimizes CPU-GPU data transfers. A cache-hit detector, based on recoded ID ranges, enables cache-aware sampling kernels to differentiate cached data from that requiring host fetches. 

On the device side, each GPU incorporates a Super-batch Degree Classifier and a Grouping-based Parallel Sampler. GARDEN introduces super-batching, which coalesces multiple mini-batches into a single contiguous super-batch to amortize setup overheads. The Super-batch Classifier performs node deduplication and classification via hash-based mapping and parallel prefix sums, avoiding costly atomics. It categorizes nodes into full nodes (requiring complete neighbor sampling) and random nodes (using probabilistic sampling), generating metadata pointers for downstream processing. The Grouping Sampler then invokes two specialized kernels: for random nodes, a Grouping-based Parallel Sampler that organizes warp threads and preloads adjacency lists into shared memory, transforming irregular global accesses into coalesced on-chip operations; for full nodes, a lightweight full-sampling kernel. These kernels run concurrently using GPU streams to maximize compute and memory utilization. Overall, This design ensures end-to-end training efficiency by aligning cache management with sampling parallelism, addressing key bottlenecks in large-scale GNN training.

\subsection{Adaptive High-Fidelity Cache Allocator}

\zhulatest{Existing static-cache designs rely on offline pre-sampling to estimate long-term access frequencies. This temporal scope is useful for stable global reuse, but extending the preprocessing period does not guarantee continuously increasing cache coverage. The following sensitivity result illustrates this limitation before we introduce AHFCA's adaptive cache organization.}

\begin{table}[!ht]
  \centering
  \scriptsize
  \setlength{\tabcolsep}{3pt}
  \caption{\zhulatest{Feature-cache hit rate of DUCATI on UK under different numbers of offline pre-sampling epochs.}}
  \label{tab:static-preepoch}
  {\color{darkgreen}
  \begin{tabular}{lccccc}
    \toprule
    Pre-epoch & 2 & 3 & 4 & 5 & 6 \\
    \midrule
    Nfeat cache hit rate (\%) & 71.73 & 75.63 & 76.88 & 75.31 & 72.62 \\
    \bottomrule
  \end{tabular}}
\end{table}

\zhulatest{On UK, the feature-cache hit rate increases from 71.73\% at two pre-epochs to 76.88\% at four pre-epochs, but decreases to 72.62\% at six pre-epochs. An excessively long offline pre-processing phase can accumulate transient access variations in the global profile, blur the persistent hot set, and reduce the accuracy of distinguishing long-term hot nodes from short-lived accesses. This non-monotonic trend shows that a longer static warm-up cannot by itself guarantee sustained hit-rate improvement. It motivates a bounded Dynamic-Cache that complements, rather than replaces, the static cache by targeting recently recurring residual hot nodes.}

Efficient caching strategies are critical for accelerating GNN training on large-scale graphs, particularly in the context of mini-batch sampling. Traditional methods often suffer from two key limitations: inaccurate offline prediction and diminishing returns from online cache maintenance. To overcome these challenges, we propose the Adaptive High-Fidelity Cache Allocator (AHFCA), a dual-phase (offline + online) caching strategy designed to enhance cache efficiency and mitigate data I/O bottlenecks during GNN training on massive graphs.

\begin{figure}
  \centering
  \includegraphics[width=\linewidth]{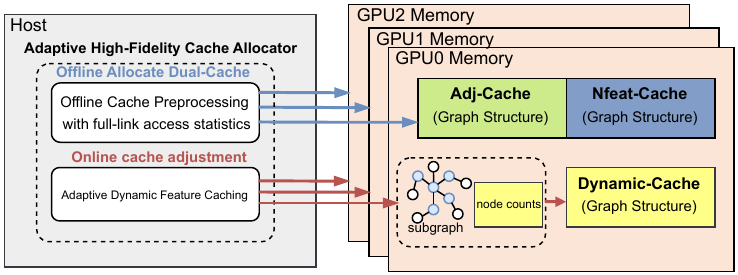}
  \caption{The basic structure of Adaptive Dynamic Feature Caching}
  \label{fig:dualcache}
\end{figure}

As illustrated in Figure~\ref{fig:dualcache}, AHFCA significantly improves the initial cache hit rate by leveraging detailed, full-link access statistics during the offline pre-sampling phase. During this phase, AHFCA allocates three distinct types of caches: adjacency cache, feature cache, and Dynamic-Cache, across all GPUs, ensuring synchronized cache availability. This strategy minimizes redundant data transfers between the host and GPU by enabling cache-aware sampling, which determines whether data is already present in the cache or needs to be fetched from main memory. Once training commences, AHFCA introduces the Adaptive Dynamic Feature Caching (ADFC) mechanism, which continuously monitors node access patterns to update the Dynamic-Cache. Using a marginal-gain model, ADFCA terminates cache updates when further adjustments yield negligible performance improvements, effectively balancing caching overhead with training efficiency.

\zhulatest{In AHFCA, the Dynamic-Cache adapts to variation in the training access sequence rather than changes in graph topology or graph structure, and it does not alter the graph sampling procedure. The static cache captures stable long-term global hotness from the offline profile, whereas the Dynamic-Cache is a bounded secondary compensation cache that supplements static-cache misses with recently recurring residual hot nodes. Because enlarging the static cache or extending its warm-up cannot avoid diminishing marginal returns, the Dynamic-Cache uses limited remaining memory to target recent accesses with higher short-term utility.}

\begin{figure}[h]
  \centering
  \includegraphics[width=\linewidth]{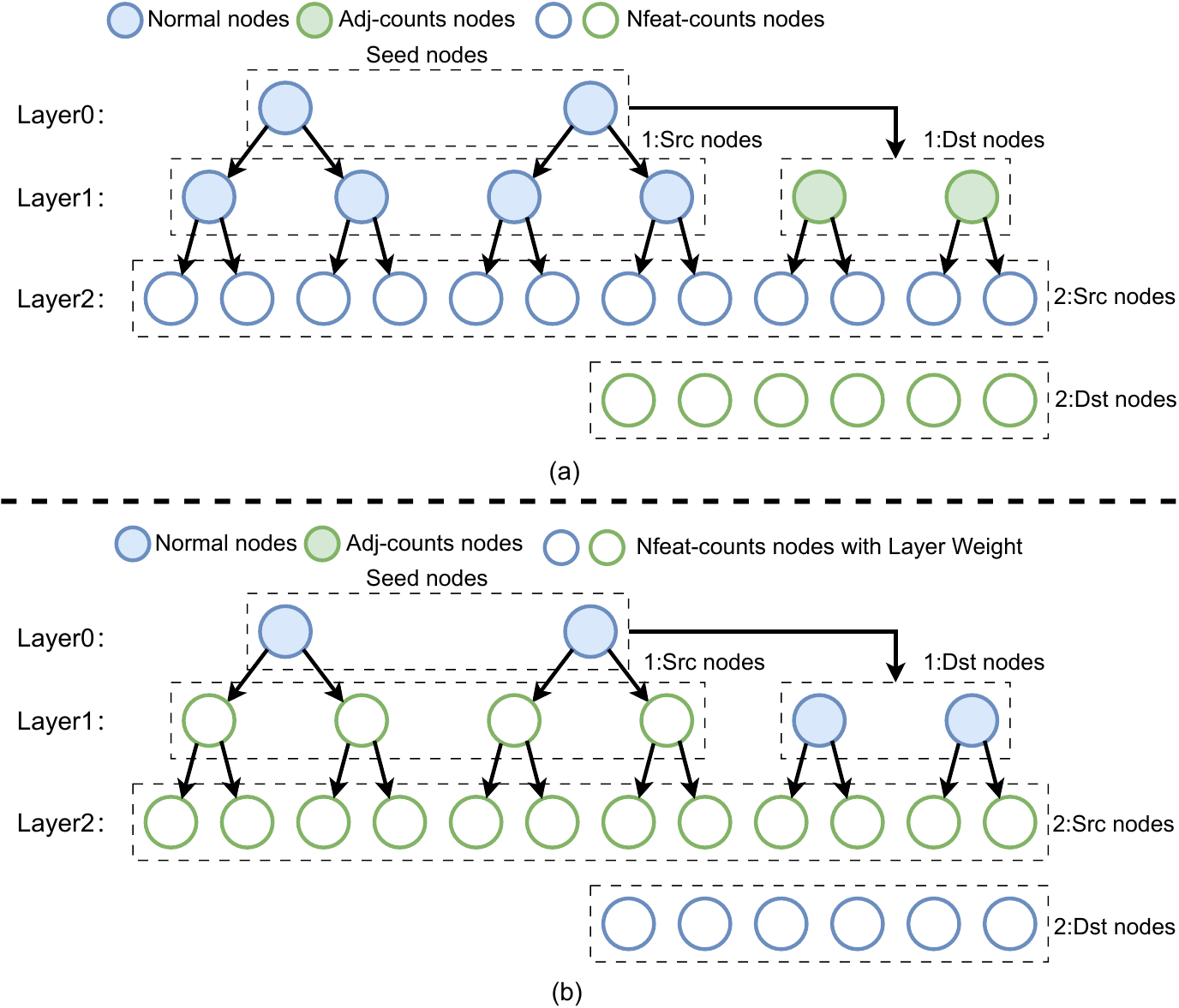}
  \caption{An example to show Node statistics during offline preprocessing: (a) conventional access frequency calculation; (b) our full-link access statistics method for offline cache preprocessing.}
  \label{fig:offlinecache}
\end{figure}

Existing offline cache preprocessing techniques often fail to accurately capture real-world GNN access patterns, leading to suboptimal cache allocation and reduced training efficiency. For example, as shown in Figure~\ref{fig:offlinecache}a, traditional approaches are limited by two significant design flaws. First, they record adjacency accesses only for destination (dst) nodes, neglecting the role of source (src) nodes, which are frequently accessed during neighbor sampling and layer-wise aggregation. Second, these methods only account for input-layer nodes in feature accesses, ignoring intermediate nodes that serve as both src and dst across multiple computational blocks, thereby underestimating inter-layer reuse.



To address these issues, we propose a more comprehensive approach that captures full-link access patterns, including all src, dst, and seed nodes across every layer of the GNN. By adopting a layer-aware weighting strategy, we account for the reuse intensity of nodes based on their sampling depth. The refined feature access model, denoted as:
\footnotesize
\begin{equation}
W_{feat}(v) = \mathbbm{1}(v \in V_{in}) + \sum_{l=1}^{L} \omega_l \cdot \mathbbm{1}(v \in V_{src}^{(l)}) + \mathbbm{1}(v \in V_{dst}) 
\end{equation}
\normalsize
allows us to accurately track the contributions of each node to the overall access frequency, ensuring more reliable cache prioritization and improved hit rates.

\zhulatest{The static cache is constructed from full-link offline access statistics collected during the offline profiling period. This profiling period is intended to estimate stable long-term global hotness across the training workload, rather than short-term fluctuations.}
\par\smallskip
\zhulatest{Because the static cache optimizes global frequency, simply increasing its capacity or extending the profiling period cannot indefinitely improve the hit rate; once the most reusable global items have been selected, the marginal benefit of additional entries decreases. AHFCA therefore uses the Dynamic-Cache as a bounded secondary compensation cache, not as an independent replacement for the static cache. It targets only residual feature accesses that are missed by the static cache and recur within recent access windows.}
\par\smallskip
\zhulatest{Although high-fidelity offline caching improves initial performance, the long-term profile does not fully describe every recent training access sequence. To address this limited residual locality, AHFCA incorporates the ADFC mechanism. ADFC uses an Exponentially Weighted Moving Average (EWMA) to emphasize recent access frequency. The Dynamic-Cache $\mathcal{C}_{\text{dyn}}$ is updated periodically based on these hotness scores, as represented by:}
\color{darkgreen}
\begin{equation}
    \text{Score}(v,t)=\gamma\text{Score}(v,t-1)
    +\sum_{i\in\mathcal{B}_{\text{cachehit},t}}\mathbb{I}(i=v)
\end{equation}
\color{black}
\zhulatest{Here, $\text{Score}(v,t-1)$ is the historical hotness score of node $v$, $\gamma\in(0,1)$ is the decay coefficient, and $\mathcal{B}_{\text{cachehit},t}$ denotes the cache-hit nodes in the current batch. The summation counts the occurrences of node $v$ among these cache-hit nodes. Thus, the EWMA retains a decayed history of previous hotness while assigning greater weight to recent access frequency.}
\par\smallskip

\zhulatest{At each update interval $T_{\text{update}}$, the system selects the top $K_{\text{dyn}}$ nodes with the highest hotness scores to update the Dynamic-Cache using a Top-$K$ operation. This bound prevents the compensation cache from expanding into a second full static cache and limits the memory and maintenance overhead of online adaptation:}
\begin{equation}
    \mathcal{C}_{\text{dyn}, t+1} = \text{TopK}\left( \left\{ \text{Score}(v, t) \right\}_{v \in \mathcal{V}},\, K_{\text{dyn}} \right)
\end{equation}
\zhulatest{This bounded update policy addresses the overhead of online cache maintenance. In the measured configuration, Dynamic-Cache updates are performed for a fixed schedule of 10 repeated epochs and then stop, while the cache-processing work overlaps with GPU training. Thus, AHFCA balances adaptation to recent residual hotness with bounded maintenance cost: it changes feature residency without altering the graph sampling procedure or the training objective.}
\par\smallskip
\zhulatest{Larger Dynamic-Cache capacities may provide only limited additional hit-rate improvements while increasing the maintenance cost of EWMA score updates and Top-$K$ selection. Accordingly, the Dynamic-Cache capacity is exposed as a tunable parameter and remains bounded by the GPU memory available after reserving space for the model and static caches.}
\zhulatest{The runtime, memory, and accuracy effects of this bounded Dynamic-Cache maintenance policy are evaluated in Table~\ref{tab:dyn-overhead}.}
\par\smallskip
\zhulatest{The following step-wise comparison reports the marginal diminishing-return ratio for DUCATI and GARDEN. A negative $r_i$ means that the current expansion provides a larger marginal improvement than the preceding expansion. Although local fluctuations remain, GARDEN exhibits negative ratios at multiple expansion steps and does so more frequently than DUCATI. This pattern shows that GARDEN's cache strategy effectively mitigates marginal diminishing returns by recovering useful cache benefits across multiple expansion stages.}
\begin{table}[t]
  \centering
  \footnotesize
  \setlength{\tabcolsep}{3.5pt}
  \caption{\zhulatest{Step-wise marginal diminishing-return ratios $r_i$ on the UK dataset. The two panels partition the 11 expansion steps for a compact single-column presentation.}}
  \label{tab:cacheratio2}
  {\color{darkgreen}
  \begin{tabular}{@{}crr@{\hspace{1.4em}}crr@{}}
    \toprule
    \multicolumn{3}{c}{\textbf{Steps 1--6}} & \multicolumn{3}{c}{\textbf{Steps 7--11}} \\
    \cmidrule(lr){1-3}\cmidrule(lr){4-6}
    Step & DUCATI & GARDEN & Step & DUCATI & GARDEN \\
    \midrule
    1  & 0.46 & -0.15 & 7  & 0.15 & -0.30 \\
    2  & 0.46 & 0.56  & 8  & 0.71 & 0.69 \\
    3  & 0.40 & -0.73 & 9  & 0.05 & -2.21 \\
    4  & -2.50 & -0.01 & 10 & 0.53 & 0.66 \\
    5  & 0.16 & 0.44  & 11 & 1.96 & 0.03 \\
    6  & 0.17 & 0.23  &    &      &      \\
    \bottomrule
  \end{tabular}}
\end{table}

\subsection{Grouping-based Parallel Sampler}

In billion-scale GNN training, the sampling phase often faces critical system-level bottlenecks. These arise from the mismatch between the irregular topology of real-world graphs and the architectural characteristics of GPU hardware. Specifically, the adjacency matrix representation of large-scale graphs consumes substantial memory resources. During sampling, random accesses to adjacency lists exhibit high stride variability, which reduces spatial locality, particularly when high-degree nodes are involved. These irregular memory access patterns hinder the coalescing of memory transactions, resulting in suboptimal memory bandwidth utilization and decreased sampling kernel efficiency.

Moreover, thread load imbalance and suboptimal hardware utilization further impede performance. In mini-batch training, heterogeneity in node degrees introduces significant variance in workload distribution across computing units. This imbalance, exacerbated by the irregular branching structure of graphs and conservative fanout hyperparameters, impairs the ability of sampling kernels to maintain high thread occupancy within GPU streaming multiprocessor (SM) warps. Consequently, warp divergence occurs, leading to idle threads and reduced parallel execution efficiency. Thus, the primary challenges are: \textit{(1) high computational overhead due to irregular memory accesses during sampling, (2) dynamic load imbalance, and (3) the mismatch between sampling strategies and GPU hardware architectures.}

To address these challenges, we propose a fine-grained node classification and sampling optimization framework built upon DUCATI's offline adjacency cache preprocessing. DUCATI’s offline graph reordering ranks nodes based on historical access frequency, assigning smaller identifiers to high-frequency nodes. This reordering establishes a solid foundation for load balancing and optimization within the sampling kernels. During the preprocessing stage, the system determines the optimal adjacency list cache capacity, defines cache boundaries, and selectively caches adjacency information for nodes with identifiers below a designated threshold. This comprehensive preprocessing improves memory locality, workload distribution, and resource utilization during the sampling phase.

Building on this foundation, we implement a fine-grained classification mechanism for seed nodes based on their dynamic characteristics and structural properties. Specifically, we use the sampling hyperparameter, fanout, to categorize nodes into two types: random nodes (with degrees greater than the fanout threshold) and full nodes (with degrees less than or equal to the fanout threshold). For random nodes, the sampling kernel generates random offsets to select neighbors, while for full nodes, all neighbors are directly sampled in one operation.


\begin{figure}[h]
  \centering
  \includegraphics[width=\linewidth]{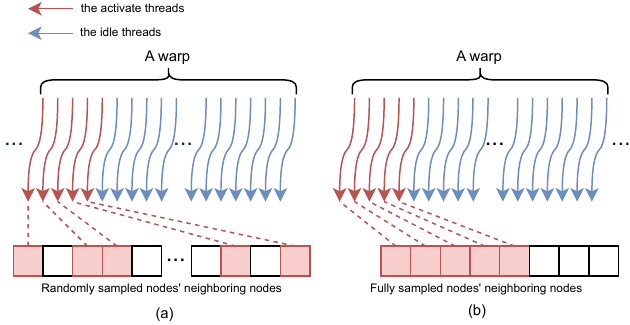}
  \caption{Sampling processes for (a) random nodes and (b) full nodes in conventional sampling systems.}
  \label{fig:sampler}
  \vspace{-10pt}
\end{figure}


As illustrated in \autoref{fig:sampler}a-b, most sampling systems allocate GPU resources at the warp or block level, dedicating computational units to either random or full nodes. However, this approach often leads to low thread utilization, especially when processing low-degree nodes or using small fanout values. With many threads in a warp remaining idle, hardware resources are underutilized. Additionally, for random nodes, the irregular memory accesses hinder memory transaction coalescing, reducing data throughput and wasting bandwidth.

\begin{figure}[h]
  \centering
  \includegraphics[width=\linewidth]{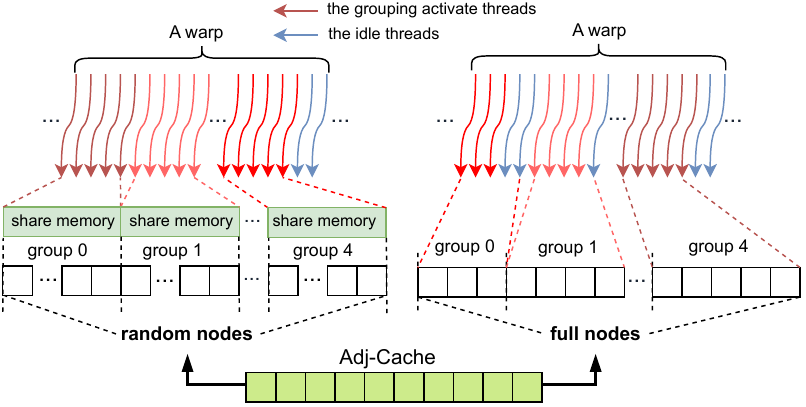}
  \caption{Grouping Parallel Sampler operation: with fanout=5, a single warp simultaneously processes six nodes.}
  \label{fig:Groupsampler}
\end{figure}

To optimize thread load balancing and execution efficiency, we introduce grouping and scheduling optimizations for these two types of nodes. During sampling preprocessing, we rely on precomputed node degrees to facilitate efficient parallel sampling computations. \zhu{The complete sampler method is shown in \autoref{fig:Groupsampler}}, for random nodes, the Grouping Parallel Sampler partitions 32 threads within a warp into multiple groups, where the width $g$ of each group corresponds to the vertex sampling fanout. This mapping allows a single warp to process $\lfloor 32/g \rfloor$ vertices in parallel. As threads within each group execute identical logic (selecting target nodes and generating random offsets), warp divergence is eliminated, resulting in synchronized execution and higher thread occupancy. 

In contrast, full nodes do not require random selection. Instead, they read the complete neighbor list of each node, which results in contiguous memory access, eliminating the need for shared memory allocation. The grouping method for full nodes is similar to that of random nodes, ensuring efficient memory utilization.
To efficiently manage both types of sampling, we schedule the random and full sampling kernels to execute concurrently through stream processing, maximizing hardware resource utilization and improving system throughput. 

For memory access optimization, the system allocates contiguous, exclusive buffers in shared memory for nodes with degrees below a preset threshold $D_{max}$, with a buffer size set to $D_{max}$. For example, setting $D_{max}$ to 64 for the Papers100M dataset covers about 97\% of nodes. This configuration ensures that the majority of nodes benefit from optimized random access while managing shared memory resources effectively.

To facilitate efficient population of these buffers, the system employs a Warp-based Collaborative Coalesced Loading strategy. Upon kernel launch, the warp serves as the execution unit for neighbor retrieval, with threads collaboratively fetching adjacency lists into shared memory via vectorized instructions and strided access patterns. This transforms fragmented random accesses into efficient, coalesced transactions, thereby maximizing memory bandwidth utilization. Once the adjacency data is staged in shared memory, subsequent random sampling operations by warp threads proceed with register-level latency.
For nodes with degrees exceeding the $D_{max}$ threshold, the framework employs a direct-access fallback strategy. Instead of using shared memory, threads perform random indexed lookups directly in global memory to retrieve the required neighbor information. Although this incurs higher latency, the relatively low frequency of high-degree nodes preserves overall bandwidth and efficiency.

By integrating the Adj-Cache and Unified Virtual Addressing (UVA)\cite{UVA}, the system ensures a robust memory access strategy throughout the sampling process. Initially, adjacency data is fetched from the Adj-Cache. If a cache miss occurs, the system seamlessly uses UVA to retrieve the data from host memory. This unified caching and retrieval framework ensures high data throughput and system robustness, even when working with billion-scale graphs. As shown in \autoref{fig:Groupsampler}, thread groups use a coalesced load mechanism to transfer adjacency lists from global memory to shared memory in one operation. The shared memory structure, with its 32-way bank parallelism, allows low-latency access, ensuring that subsequent neighbor accesses are efficiently executed within shared memory.
Our grouping-based parallel sampler further improves stochastic neighbor access by reorganizing thread-group execution to increase warp occupancy and reduce branch divergence under the skewed degree distributions common in power-law graphs. Importantly, it preserves the original fixed-fanout uniform sampling protocol used in standard frameworks such as GraphSAGE~\cite{GraphSAGE}: we do not alter sampling probabilities, apply pruning or reweighting, or otherwise change the sampling distribution. As a result, these scheduling optimizations preserve the unbiasedness of the stochastic gradient estimator and maintain training fidelity, as also demonstrated in our empirical results (Section~\ref{subsec:Accuracy}).

This memory optimization strategy operates in conjunction with the overall caching system. First, high-frequency adjacency lists are preloaded into GPU caches to reduce CPU-GPU data migration overhead. Second, shared memory buffering is activated for nodes meeting the degree criteria, improving memory locality during sampling. This two-tiered caching design, integrating global and shared memory caches, significantly reduces memory access latency and enhances bandwidth utilization during GNN training.

\subsection{Super-batch Degree Classifier}
In grouping-based parallel sampling, pre-sampling node classification is essential for optimizing memory access and load balance. However, existing approaches typically perform classification independently for each mini-batch, incurring substantial redundant computation and synchronization overhead. In GPU environments, accurate node categorization often relies on atomic operations to ensure correctness, which leads to severe thread contention and increased latency. Moreover, graph learning workloads commonly employ small mini-batches, whose limited workload fails to fully occupy GPU resources, resulting in low SM utilization. Prior work (e.g., gSampler~\cite{2023gSampler}) has shown that increasing batch size improves GPU throughput. However, under the conventional mini-batch paradigm, this also amplifies the cost of repeatedly executing classification and deduplication logic.

\begin{figure}[h]
  \centering
  \includegraphics[width=\linewidth]{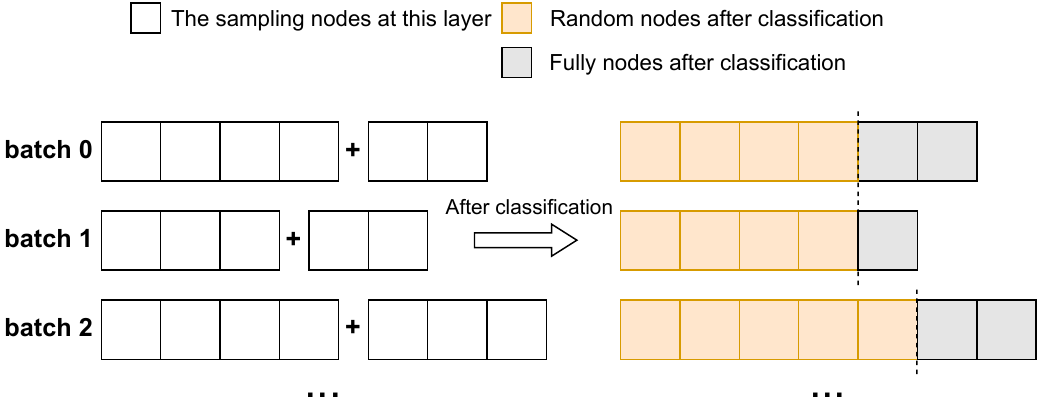}
  \caption{\zhu{Super-batch classification: nodes are categorized as random or full nodes with per-batch segmentation before sampling.}}
  \label{fig:superbatch}
\end{figure}

\textbf{Super-batch Construction and Deduplication.} To address these limitations, we propose a Super-batch Degree Classifier that jointly optimizes batching, deduplication, and node classification. The key idea is to aggregate multiple mini-batches into a single super-batch with contiguous storage, enabling batch-wide preprocessing and classification in one pass. A batch pointer, denoted as \textit{BatchPtr}, encodes sub-batch boundaries, preserving logical isolation between mini-batches while allowing unified processing.

During super-batch construction, node sets at each sampling layer are incrementally merged with previously sampled nodes. As illustrated in \autoref{fig:superbatch}, this process eliminates redundancy both within and across mini-batches. We maintain a GPU-resident hash table that maps global node identifiers to unique entries, ensuring that each node appears exactly once in the super-batch view. This design avoids repeated node traversals and enables efficient deduplication via parallel hash lookups.

\textbf{Degree-based Classification without Atomics.} On top of the deduplicated super-batch, we introduce a lightweight degree-based classification scheme. Nodes are categorized into two types:
Full Nodes, whose degrees do not exceed the fanout and are sampled exhaustively; and
Random Nodes, which require stochastic neighbor selection. The classification result is encoded as a logical partition boundary, \textit{BoundPtr}, which separates full nodes and random nodes within each sub-batch. Rather than relying on atomic counters, the classifier transforms category assignment into deterministic index computation.

\begin{algorithm}[h]
    \scriptsize
    \renewcommand{\algorithmicrequire}{\textbf{Input:}}
    \renewcommand{\algorithmicensure}{\textbf{Output:}}
    \caption{The Super-batch Degree Classifier Algorithm}
    \label{alg:classifier}
    \algrenewcommand\alglinenumber[1]{\makebox[1.8em][l]{#1:}}
    \begin{algorithmic}[1]
        \Require $\text{MergeTensor}[0\ldots N-1]$, $\text{BatchPtr}[0\ldots K]$, $\text{indptr}$, $\text{fanout}$
        \Ensure $\text{UniqueClassify}$, $\text{BoundPtr}$, $\text{DegreeFlag}$

        \Function{HashmapsForUniqueClassify}{\textit{MergeTensor}, \textit{fanout}}
            \State $\text{UniqueTensor} \gets \Call{UniqueHashmaps}{\textit{MergeTensor}}$
            \For{each thread $tid = 0 \ldots N-1$ \textbf{in parallel}}
                \State $v \gets \text{UniqueTensor}[tid]$; $deg \gets \text{indptr}[v+1] - \text{indptr}[v]$
                \State $\text{DegreeFlag}[tid] \gets (deg \le \text{fanout}) ? 1 : 0$
            \EndFor
            \State \Return $\text{DegreeFlag}$
        \EndFunction

        \Function{BatchPartition}{\textit{UniqueTensor}, \textit{DegreeFlag}, \textit{BatchPtr}}
            \State $\text{DegreeScan} \gets \Call{exclusiveSum}{\textit{DegreeFlag}}$
            \For{$i = 0 \ldots K-1$} \Comment{Boundary per batch}
                \State $\text{BoundPtr}[i] \gets \text{DegreeScan}[\text{BatchPtr}[i+1]] - \text{DegreeScan}[\text{BatchPtr}[i]]$
            \EndFor
            \For{each thread $tid = 0 \ldots N-1$ \textbf{in parallel}}
                \State Find $i$ such that $\text{BatchPtr}[i] \le tid < \text{BatchPtr}[i+1]$
                \State $batchStart \gets \text{BatchPtr}[i]$; $batchID \gets tid - batchStart$
                \State $pos \gets \text{DegreeScan}[tid] - \text{DegreeScan}[batchStart]$
                \If{$\text{DegreeFlag}[tid] == 1$} \Comment{Full node}
                    \State $outIdx \gets batchStart + pos$
                \Else \Comment{Random node}
                    \State $Rank \gets batchID - pos$; $outIdx \gets \text{BoundPtr}[i] + Rank$
                \EndIf
                \State $\text{UniqueClassify}[outIdx] \gets \text{UniqueTensor}[tid]$
            \EndFor
            \State \Return $\text{UniqueClassify}, \text{BoundPtr}$
        \EndFunction
    \end{algorithmic}
\end{algorithm}

As shown in Algorithm~\ref{alg:classifier}, the \texttt{HashmapsForUniqueClassify} kernel simultaneously performs node deduplication and category marking (lines 2--8). Each node is assigned a binary flag stored in the \textit{DegreeFlag} tensor, where 1 denotes a full node and 0 denotes a random node (line 5). Subsequently, the \texttt{BatchPartition} kernel applies a parallel prefix-sum over \textit{DegreeFlag} using CUB’s \texttt{ExclusiveSum}\cite{cub} (line 10), producing the \textit{DegreeScan} tensor. This scan yields, for each node, its offset within the full-node partition of its corresponding sub-batch. The number of full nodes per sub-batch is then computed using a lightweight difference operation over scan results, yielding \textit{BoundPtr} (line 12). Importantly, this approach converts classification into spatial placement: each thread independently computes its write position in the output tensor (lines 15--22), completely eliminating atomic operations and synchronization stalls.

\textbf{Cache-friendly Layout and Sampling Readiness.} Using the computed offsets, full nodes are placed contiguously at the beginning of each sub-batch, followed by random nodes. For random nodes, their position is derived as \textit{Rank} = \textit{batchID} - \textit{pos}, where \textit{pos} is the number of preceding full nodes obtained from the prefix-sum. This layout guarantees contiguous storage for both node types within each sub-batch, significantly improving memory locality for downstream sampling kernels.
Crucially, this design collapses preprocessing overhead from being proportional to the number of mini-batches to a single batch-wide operation. Once \textit{UniqueClassify} and \textit{BoundPtr} are generated, the sampler can directly locate target nodes through simple pointer arithmetic, enabling highly efficient differentiated sampling.


Overall, the Super-batch Degree Classifier achieves efficiency gains through a three-level optimization: (1) Algorithmically, it fuses deduplication, classification, and layout into a single super-batch pass, eliminating redundant work; (2) At the parallel execution level, it replaces atomic operations with prefix-sum–based index computation, removing thread contention; (3) At the hardware utilization level, it increases effective batch size, improving SM occupancy and fully exploiting GPU parallelism. Together, these optimizations enable scalable, cache-friendly, and synchronization-free sampling for billion-scale GNN training.

\section{Implementation}

We implement GARDEN atop PyTorch\cite{pytorch}, leveraging DGL v1.0.0 for efficient graph loading and subgraph sampling via its data loader module. For the Adaptive High-Fidelity Cache Allocator, we introduce two Python classes: \texttt{Constructor} for building the dual-cache structures (Adj-Cache, Nfeat-Cache, and dynamic cache) and \texttt{Allocator} for runtime allocation and updates. When GPU memory is underutilized, GARDEN repurposes idle VRAM as an additional cache layer to further reduce I/O latency. The Grouping Parallel Sampler is realized through custom CUDA kernels that enforce memory-aware access patterns, including warp-based collaborative loading and shared memory buffering. Following gSampler's matrix-style API, we encapsulate these kernels into modular interfaces, enabling easy integration with diverse GNN models. For super-batching, we extend gSampler's CUDA implementation with the Super-batch Degree Classifier, optimizing node deduplication and partitioning before seamless integration into DGL's sampling pipeline. Multi-GPU training on a single machine uses NCCL\cite{nccl} for gradient synchronization, supporting data-parallel execution. GARDEN's modular design allows individual components to be incorporated into other frameworks with minimal code changes, promoting reusability.

\section{Evaluation}
\subsection{Experiment Setup}


\begin{figure*}[ht]
  \centering
  \includegraphics[width=\textwidth]{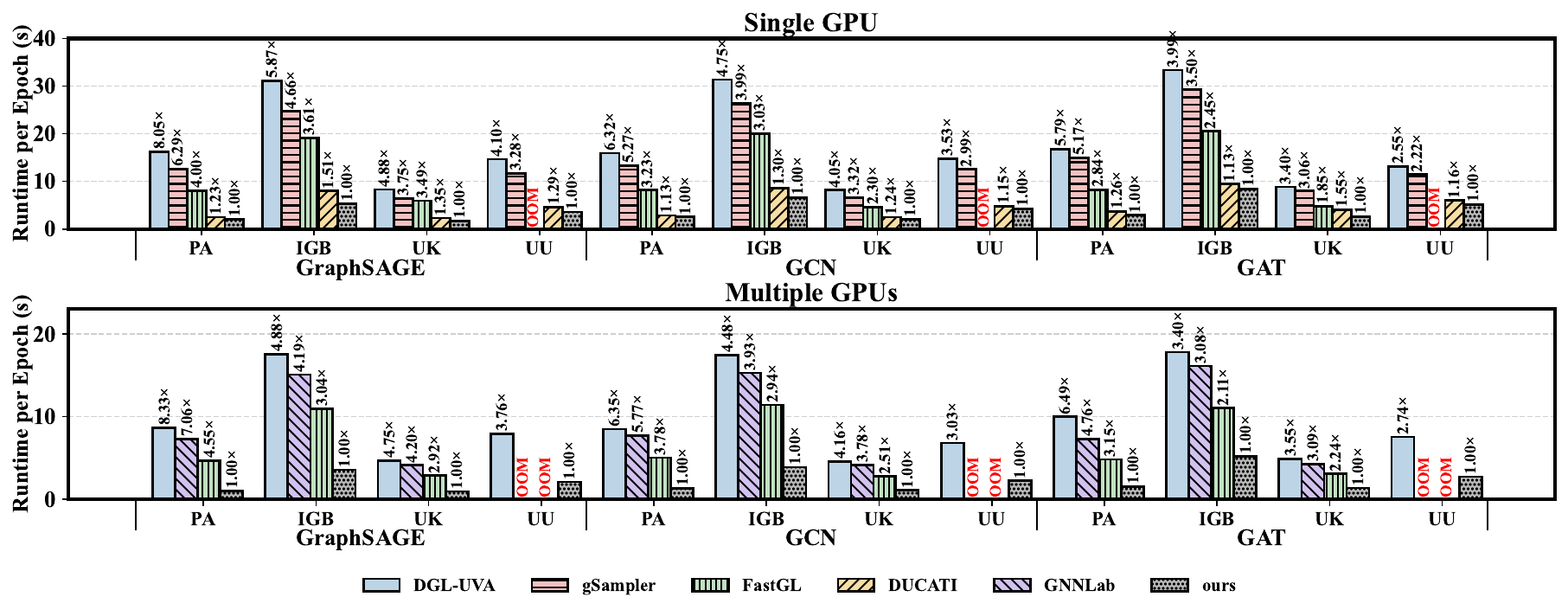}
  \caption{\zhulatest{Average runtime per epoch of three GNN models across multiple datasets under single- and multi-GPU configurations. Numbers above the bars are baseline-to-GARDEN runtime ratios, and OOM denotes a GPU-memory capacity failure under the reported configuration.}}
  \label{fig:e2e}
\end{figure*}

{\sloppy
\textbf{Benchmarks.} We evaluate GARDEN using three representative GNN models: GraphSAGE~\cite{GraphSAGE}, Graph Convolutional Network (GCN)~\cite{gcn}, and Graph Attention Network (GAT)~\cite{gat}. All models employ 3-hop random neighbor sampling with fanouts of [5, 10, 15] for the first three layers, matching prior configurations. The batch size is 8000, consistent with GNNLab baselines. GraphSAGE and GCN use a hidden dimension of 128; GAT employs 8 attention heads, each with dimension 8.}
\par

\zhulatest{\textbf{Datasets.} We test on four real-world graphs of varying scales, as summarized in Table~\ref{tab:datasets}. These include Papers100M (PA, a citation network) from the Open Graph Benchmark, IGB-large (a large-scale homogeneous academic citation graph from the Illinois Graph Benchmark), and two billion-scale web graphs, UK-2006-05 (UK) and UK-Union (UU, a union of UK web snapshots). For UK and UU, which lack official node features, we use randomly generated features with length 128 for runtime and scalability evaluation; these datasets are therefore excluded from the accuracy comparison.}

\begin{table}[h]
  \centering
  \caption{The statistics of datasets used in this work (M: Million, B: Billion).}
  \label{tab:datasets}
  \begin{tabular}{lcccc}
    \toprule
    Dataset & Nodes & Edges & Features & Classes \\
    \midrule
    IGB-large\cite{igb}     & 100M       & 1.2B & 128  & 19  \\
    Papers100M\cite{2020OGB}    & 111M       & 1.6B & 128   & 172 \\
    UK-2006-05\cite{uk}    & 77.7M      & 3.0B & 128   & 100 \\
    UK-Union\cite{uk}      & 134M       & 5.5B & 128   & 100 \\
    \bottomrule
  \end{tabular}
\end{table}



\zhu{\textbf{Baselines.} We compare GARDEN against state-of-the-art systems: DGL-UVA\cite{dgl-uva}, gSampler\cite{2023gSampler}, GNNLab\cite{2022GNNLab}, DUCATI\cite{2023DUCATI}, and FastGL\cite{2024fastgl}. DGL-UVA conducts sampling using the GPU by utilizing the UVA technique. GNNLab uses efficient caching and multi-GPU parallelism for sampling and training. GSampler provides a general GPU-based sampling framework. DUCATI extends GNNLab with dual caching to reduce I/O. FastGL combines match-reorder and memory-aware techniques for large-scale sampling.}

\zhulatest{PaGraph\cite{PaGraph} and AliGraph\cite{AliGraph2019} use CPU-based sampling and graph-processing paths in their original implementations. Because these paths differ from our GPU-accelerated sampling and caching pipeline, we select systems with comparable GPU-accelerated execution paths that can be evaluated under matched datasets, sampling configurations, hardware, and timing protocols as direct baselines.}

\zhu{Notably, several baselines such as FastGL require the entire graph adjacency information to be stored in GPU memory, which is infeasible for certain billion-scale datasets. During the overall training process, we follow the convention adopted by DUCATI and FastGL and exclude the offline preprocessing time from the end-to-end training time. As demonstrated in prior work\cite{offlinebaseline}, in practical training scenarios, models typically require on the order of thousands of epochs; consequently, the preprocessing time is negligible compared to the total training time.}

\textbf{Environment.} Experiments run on a server with two AMD EPYC 7702 CPUs (80 cores total), 512 GB DRAM, and six NVIDIA RTX 3090 GPUs (24 GB VRAM each). The software environment includes Python v3.8.10, PyTorch v1.12.1, CUDA v11.6, and DGL v1.0.0.

\subsection{Overall Performance}
\zhulatest{As illustrated in \autoref{fig:e2e}, we report average end-to-end runtime per epoch for GARDEN and the baselines across the evaluated datasets and GNN models. The experiments include both single-GPU and multi-GPU configurations. The bar heights show absolute runtime, while the labels above the bars show the corresponding baseline-to-GARDEN runtime ratios. An OOM entry indicates that the workload exceeds the available GPU memory under the reported configuration.}

\zhulatest{The absolute runtimes and per-case ratios show the combined effect of GARDEN's cache, sampling, and batching optimizations. Across the evaluated workloads, GARDEN achieves average end-to-end training speedups of 4.54$\times$, 3.95$\times$, 4.04$\times$, 1.27$\times$, and 2.97$\times$ over DGL-UVA, gSampler, GNNLab, DUCATI, and FastGL, respectively. The largest gap is observed against DGL-UVA, whose stochastic sampling generates scattered feature requests and reduces effective memory-bandwidth utilization. GARDEN's warp-centric, coalesced data loader restores more efficient memory access and achieves up to a 4.5$\times$ speedup in training throughput.}

\zhulatest{Compared with DUCATI, GARDEN provides an additional 1.13--1.55$\times$ improvement and reduces sampling latency by 27--32\%. These gains result from coordinating the Grouping-based Parallel Sampler with the Super-batch Degree Classifier: the former transforms irregular accesses into coalesced shared-memory operations, while the latter reduces redundant vertex expansion and amortizes synchronization across a super-batch. Together with the cache design, these optimizations balance data movement, sampling, and batch-level execution.}

\subsection{Performance Breakdown}
To precisely attribute the performance gains and validate the effectiveness of our three proposed techniques, we conduct detailed component-level analyses comparing GARDEN against state-of-the-art baselines across multiple datasets and configurations.

\begin{figure}[h]
  \centering
  \vspace{-10pt} 
  
  \includegraphics[width=\linewidth]{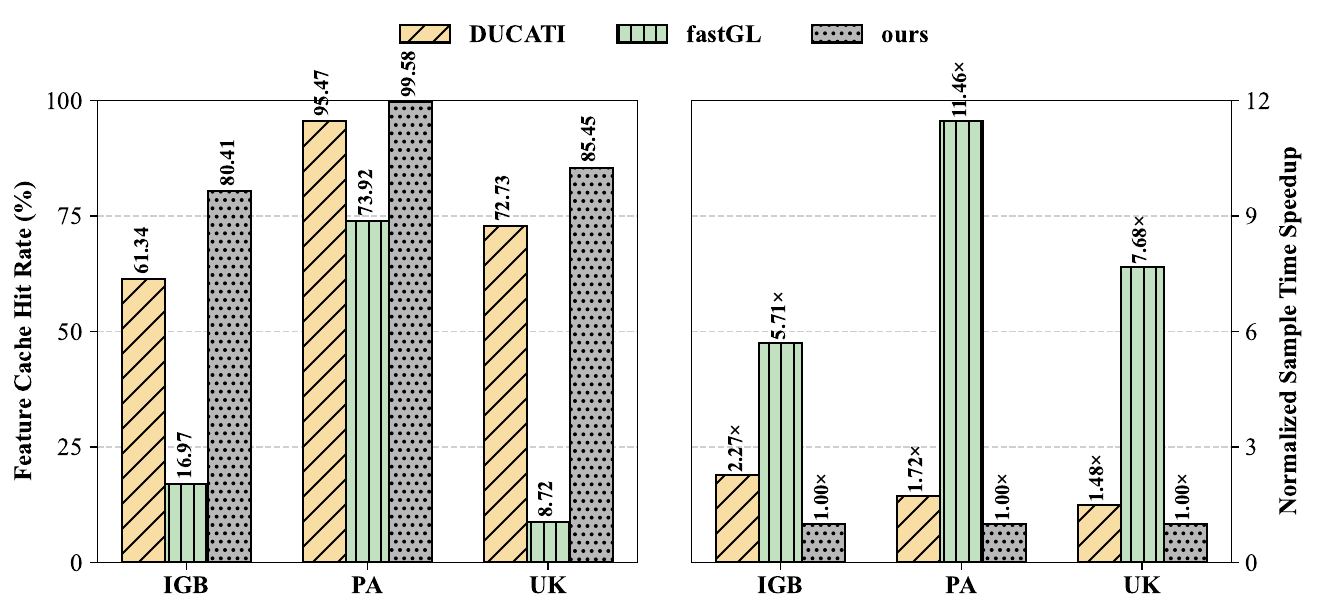}
  
  \vspace{-6pt} 
  {\small (a) \hspace{0.4\linewidth} (b)} 
  
  \vspace{-8pt}  
  \caption{Cache performance comparison: (a) feature cache hit rate, (b) node feature selection time per iteration.}
  \label{fig:exp-cache}
\end{figure}

\begin{figure}[t]
  \centering
  \begin{minipage}[t]{0.45\columnwidth}
    \vspace{0pt}
    \centering
    \includegraphics[width=\linewidth]{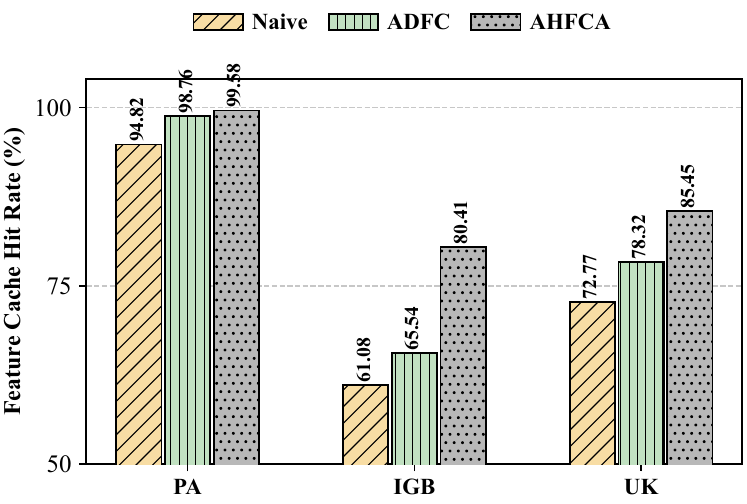}
    \captionof{figure}{\zhulatest{Feature-cache hit rate for Naive, ADFC, and AHFCA.}}
    \label{fig:cache-hit-rate}
  \end{minipage}\hfill
  \begin{minipage}[t]{0.51\columnwidth}
    \vspace{0pt}
    \centering
    \captionof{table}{\zhulatest{Dynamic-Cache overhead on PA.}}
    \label{tab:dyn-overhead}
    {\color{darkgreen}
    \footnotesize
    \setlength{\tabcolsep}{2.5pt}
    \begin{tabular}{@{}lcc@{}}
      \toprule
      Metric & \shortstack{Static-\\Only} & \shortstack{Dynamic-\\Cache} \\
      \midrule
      E2E time (s) & 2.464 & 2.015 \\
      Repeated epochs & 0 & 10 \\
      Update/epoch (s) & 0 & 2.83 \\
      Peak memory (GB) & 17.68 & 18.92 \\
      Final accuracy (\%) & 67.23 & 67.23 \\
      \bottomrule
    \end{tabular}}
  \end{minipage}
\end{figure}

\zhulatest{\textbf{Effectiveness of the Adaptive High-Fidelity Cache Allocator.} Built on the single-machine multi-GPU dual-cache runtime, where preprocessing is performed on one GPU and broadcast to others, AHFCA combines multi-dimensional profiling with precise pre-sampling statistics to enhance cache utilization and hit rates. We benchmark it against DUCATI's dual-cache allocator and FastGL's single-cache design under varying batch sizes, tracking the feature-cache hit rate and the average time spent in the nfeat-selecting process. As shown in \autoref{fig:exp-cache}, AHFCA achieves the highest feature-cache hit rate among the evaluated baselines. By profiling source and destination vertices across all GNN layers, its memory-driven allocation strategy captures both immediate one-hop demands and multi-hop reuse patterns, retaining 5--76\% more frequently accessed features in on-chip caches than DUCATI and FastGL.}

\zhulatest{As shown in \autoref{fig:exp-cache}b, AHFCA also reduces nfeat-selecting time through high-fidelity offline allocation and adaptive online caching. By leveraging full-link access statistics, it preloads frequently reused features into GPU caches, achieving hit rates above 87\% early in execution compared with below 62\% for the baselines. The ADFC mechanism further tracks hot nodes with an Always-On Exponentially Weighted Moving Average (EWMA), reducing nfeat-selecting time to less than 26\% of its initial value within 5--8 epochs. Across PA, UK, and IGB, this produces an average 5.05$\times$ speedup in nfeat-selecting time.}

\zhulatest{The cache-isolation experiment in \autoref{fig:cache-hit-rate} quantifies the contributions of the optimized static allocation and the bounded Dynamic-Cache. Naive uses the original static dual-cache policy, ADFC applies full-link profiling to optimize static-cache allocation, and AHFCA further augments ADFC with the bounded Dynamic-Cache. Averaged across PA, IGB, and UK, ADFC increases the feature-cache hit rate by an absolute 4.65\% over Naive. Adding the Dynamic-Cache provides a further absolute improvement of 7.61\%, yielding a total absolute increase of 12.26\% from Naive to AHFCA. The additional gain is smaller on PA because its static cache already provides high coverage, whereas the gain is larger on IGB and UK because the Dynamic-Cache captures recurring hot nodes that remain uncovered by the static cache.}

\zhulatest{As reported in Table~\ref{tab:dyn-overhead}, the Dynamic-Cache overhead measurement gives average E2E times of 2.464~s for Static-Only and 2.015~s for Dynamic-Cache, peak memories of 17.68~GB and 18.92~GB, and the same final accuracy of 67.23\%. The Dynamic-Cache changes feature residency and cache maintenance only; it does not change the graph sampling procedure, neighbor fanouts, sampled neighborhoods, batch construction, model architecture, or training objective. Therefore, the two configurations produce the same final accuracy. In the measured configuration, Dynamic-Cache updates are performed for a fixed schedule of 10 repeated epochs and then stop. The 2.83~s cache-processing time per repeated epoch overlaps with GPU training and should not be added serially to the E2E runtime, so the bounded maintenance work does not translate into an equivalent additional online training cost.}

\begin{figure}[t]
  \centering
  \vspace{-10pt} 
  \includegraphics[width=\linewidth]{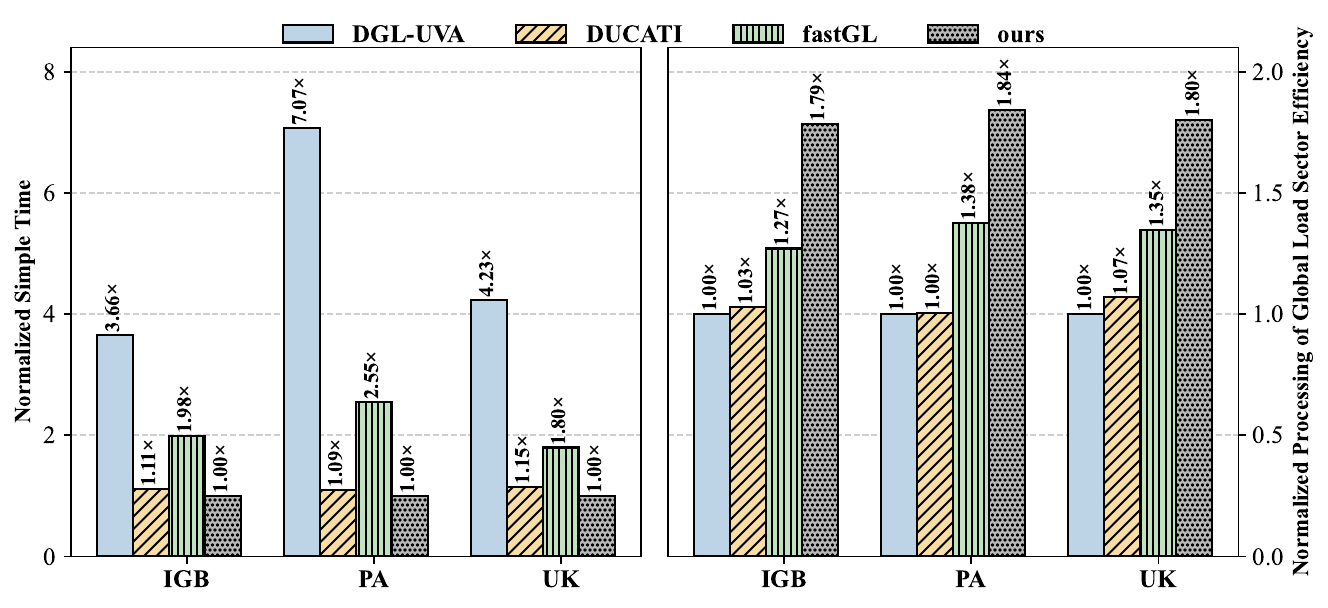}
  \vspace{-6pt} 
  {\small (a) \hspace{0.4\linewidth} (b)}
  \vspace{-1pt}  
  \caption{Comparison of sampling time and Global Load Sector Efficiency across different frameworks on the GraphSAGE model.}
  \label{fig:exp-sample}
  \vspace{-5pt}
\end{figure}

\textbf{Effectiveness of the Grouping-based Parallel Sampler.} We quantify the advantages of the Grouping-based Parallel Sampler from both micro-architectural and system-level throughput perspectives. \autoref{fig:exp-sample}a presents the end-to-end GPU kernel time for the pure sampling phase, excluding feature aggregation. Our implementation reduces sampling latency to 10–87\% of that in DGL, FastGL, and DUCATI, achieving speedups of 1.09× to 7.07×. The largest improvement occurs on the large-scale PA dataset, validating the effectiveness of our warp-level multi-node parallelism and shared-memory coalescing strategy in mitigating irregular memory access patterns.

Using Nsight Compute, we further analyze memory efficiency. Specifically, we measure Global Load Sector Efficiency (GLSE), which is the ratio of useful bytes loaded to total bytes transferred by the hardware. As shown in \autoref{fig:exp-sample}b, DGL achieves only 30\% GLSE on PA, indicating that 65\% of L2 sector bandwidth is wasted on redundant 128-byte transactions. In contrast, our Grouping Sampler buffers index accesses in shared memory, limiting random lookups to on-chip storage and issuing a single coalesced global memory load. This approach increases GLSE to 55\%, representing a 1.83× improvement over DGL.

Overall, the Grouping-based Parallel Sampler reduces compute latency and transforms irregular global memory accesses into efficient, coalesced shared-memory hits. This optimization unlocks memory-level parallelism and frees up global memory bandwidth for subsequent GNN training stages, all without requiring additional hardware.

\begin{table}[!t]
  \caption{\zhulatest{Thread execution efficiency and long-scoreboard stall for DGL-UVA and GARDEN.}}
  \label{tab:thread-balancing}
  \centering
  \scriptsize
  \setlength{\tabcolsep}{2pt}
  \renewcommand{\arraystretch}{0.92}
  {\color{darkgreen}
  \begin{tabular}{@{}lclcc@{}}
    \toprule
    Dataset & \shortstack{Mean\\degree} & Framework & \shortstack{Thread execution\\efficiency} & \shortstack{Long-scoreboard\\stall} \\
    \midrule
    \multirow{2}{*}{PA}  & \multirow{2}{*}{15.55} & DGL-UVA & 90.86\% & 97.07\% \\
                          &                         & GARDEN  & 96.75\% & 59.03\% \\
    \midrule
    \multirow{2}{*}{IGB} & \multirow{2}{*}{13.24} & DGL-UVA & 91.28\% & 93.11\% \\
                          &                         & GARDEN  & 98.22\% & 64.31\% \\
    \midrule
    \multirow{2}{*}{UK}  & \multirow{2}{*}{38.14} & DGL-UVA & 90.48\% & 92.98\% \\
                          &                         & GARDEN  & 98.85\% & 65.52\% \\
    \bottomrule
  \end{tabular}}
  \vspace{-2pt}
\end{table}

\zhulatest{As reported in Table~\ref{tab:thread-balancing}, GARDEN achieves higher thread execution efficiency and lower long-scoreboard stall than DGL-UVA on PA, IGB, and UK. The mean degree is reported to characterize and distinguish the structural properties of the evaluated graphs. The higher thread execution efficiency is attributable to GPS's degree-aware thread grouping, which keeps more execution lanes active and reduces warp divergence. The reduced long-scoreboard stall reflects lower global-memory access waiting enabled by GPS's shared-memory buffering and coalesced global-memory loading.}

\begin{figure}[h]
  \centering
  \vspace{-10pt} 
  
  \includegraphics[width=\linewidth]{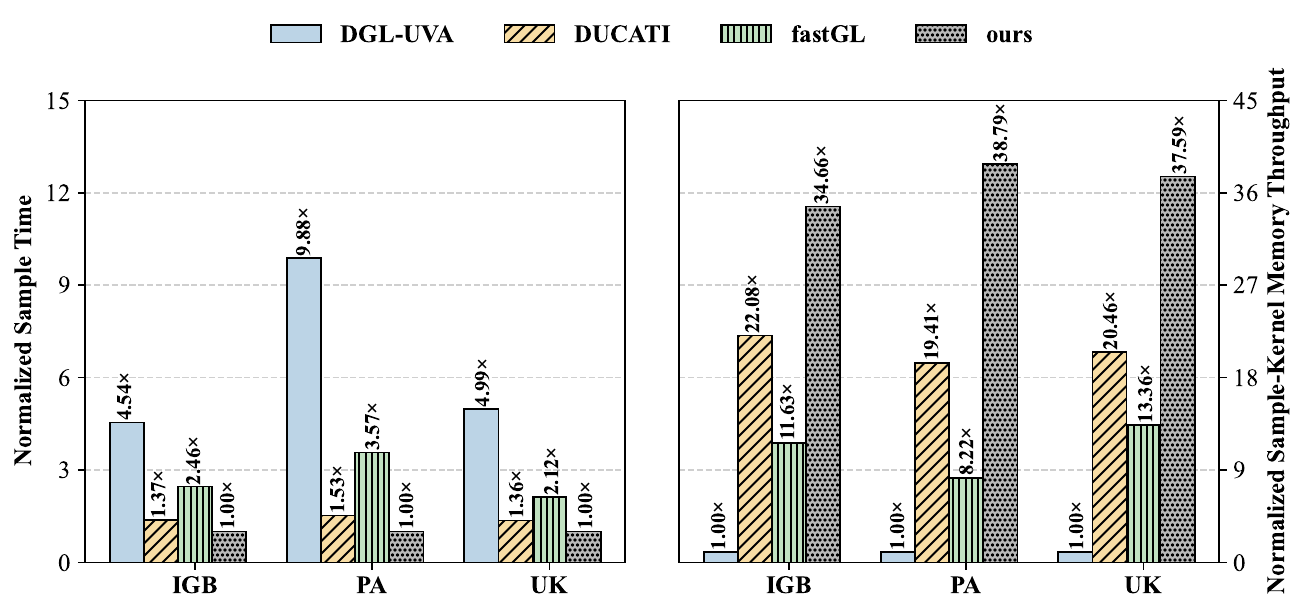}
  
  \vspace{-6pt} 
  {\small (a) \hspace{0.4\linewidth} (b)} 
  \vspace{-8pt}  
  \caption{The sampling iteration time for each epoch and memory throughput achieved by the Super-batch Degree Classifier.}
  \label{fig:exp-superbatch}
\end{figure}

\zhu{\textbf{Effectiveness of the Super-batch Degree Classifier.} The Super-batch Degree Classifier further enhances sampling efficiency and improves sampling throughput across multiple datasets. As shown in \autoref{fig:exp-superbatch}a, it achieves average sampling speedups of 6.47×, 1.42×, and 2.71× over DGL-UVA, DUCATI, and FastGL, respectively. \autoref{fig:exp-superbatch}b demonstrates substantial improvements in end-to-end throughput compared to all baseline systems, with overall speedups ranging from 8.22× to 38.79×. These gains stem from two key optimizations enabled by the Super-batch Degree Classifier: (1) it coalesces multiple mini-batches into a single super-batch, significantly improving GPU utilization; and (2) it reduces the overhead of node classification in the Grouping-based Parallel Sampler. Together, these mechanisms lead to a dramatic increase in system-level throughput.}

\begin{figure*}[!t]
  \centering
  \includegraphics[width=\textwidth]{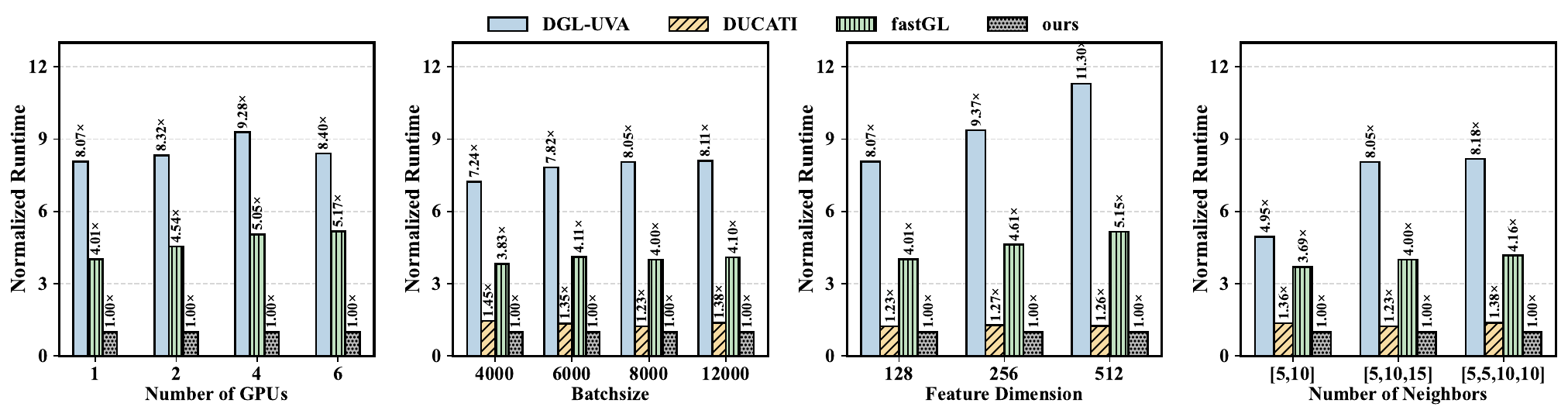}
 
  \vspace{-6pt}
  \makebox[\textwidth][c]{
    \hspace{0.04\textwidth}(a)\hspace{0.23\textwidth}(b)\hspace{0.23\textwidth}(c)\hspace{0.23\textwidth}(d)
  }
  \vspace{-15pt}
  \caption{Scalability analysis: (a) GPU count, (b) batch size, (c) feature dimension, and (d) number of sampled neighbors and GNN layers.}
  \label{fig:scalability}
\end{figure*}

\begin{figure}[!t]
  \centering
  \includegraphics[width=\linewidth]{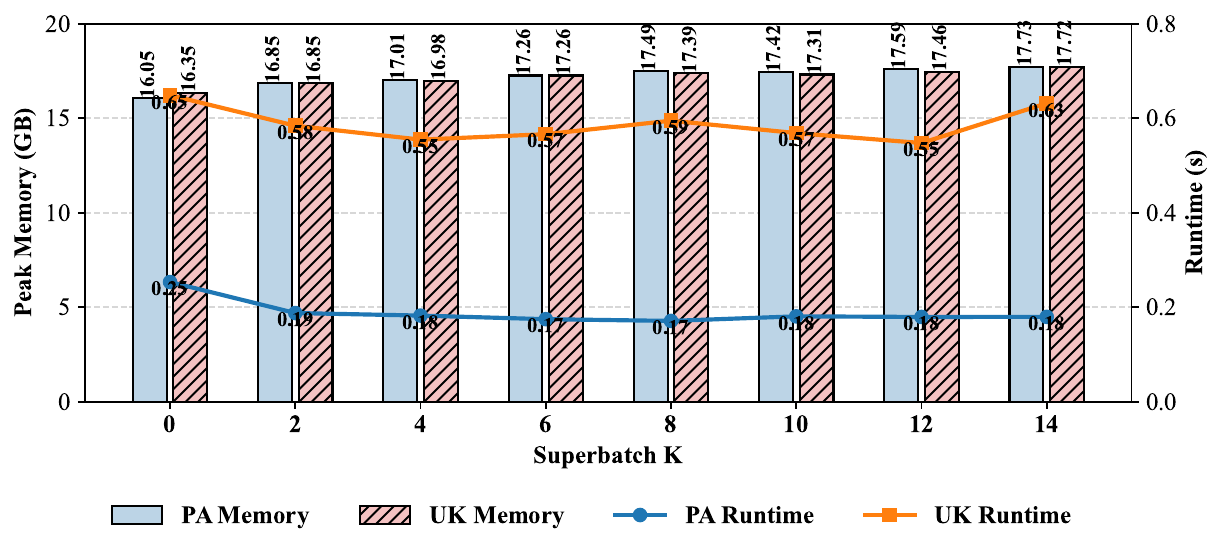}
  \caption{Impact of super-batch hyperparameter $K$ on peak GPU memory usage and sampling time per epoch across two datasets (PA and UK). The bars show peak memory consumption, while the lines represent sampling time per epoch.}
  \label{fig:superbatch-tradeoff}
\end{figure}

To quantify the trade-off introduced by super-batching, we analyze its impact on execution granularity, memory footprint, and computational cost. By aggregating multiple mini-batches into a single super-batch, GARDEN substantially increases the effective batch size. Since the number of sub-batches $K$ is typically set between 6 and 10, this design improves GPU SM occupancy and throughput, while also reducing the degree-classification overhead during sampling.

The main overheads of super-batching come from two sources: the temporary memory required by the super-batch hashmap and auxiliary tensor structures, and the cost of GPU-resident hash-based deduplication. In practice, node overlap across mini-batches is high, so the deduplication ratio is also high. For large tensors, we use pinned memory to reduce GPU memory usage, keeping the additional footprint manageable. For deduplication, we adopt a lightweight GPU hash table for parallel insertion and lookup. Although hash operations introduce extra computation relative to atomic operations in per-mini-batch classification, this cost is largely offset by eliminating repeated small-batch kernel launches and reducing atomic contention.

As shown in \autoref{fig:superbatch-tradeoff}, as $K$ increases, the sampling time of the super-batch degree classifier is 1.2-1.4$\times$ that of the naive per-mini-batch method, while peak GPU memory increases by only 6.8\%-8.5\% on average. These results indicate that the performance gains from improved parallelism and reduced synchronization overhead outweigh the modest increases in memory and hashing cost, making super-batching a favorable design for billion-scale GNN training.

\begin{figure}[!b]
  \centering
  \includegraphics[width=\linewidth]{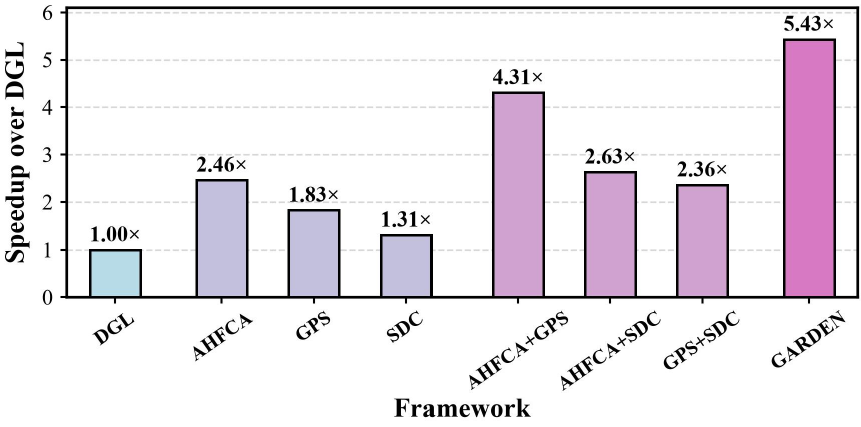}
  \caption{\zhulatest{Framework-level speedup over DGL for individual and combined GARDEN components on PA with GraphSAGE.}}
  \label{fig:exp-breakdown}
\end{figure}

\zhulatest{\textbf{Ablation Study.} On PA with GraphSAGE and DGL as the baseline, the framework-level results in \autoref{fig:exp-breakdown} report speedups of 2.46$\times$ for AHFCA, 1.83$\times$ for GPS, and 1.31$\times$ for SDC. The combined configurations achieve 4.31$\times$ for AHFCA+GPS, 2.63$\times$ for AHFCA+SDC, and 2.36$\times$ for GPS+SDC, while the complete GARDEN configuration reaches 5.43$\times$. The AHFCA+SDC configuration provides only a modest gain over AHFCA alone, whereas SDC provides a more evident improvement when combined with GPS: GPS+SDC reaches 2.36$\times$, compared with 1.83$\times$ for GPS alone. This behavior arises because SDC's degree-classification results are directly consumed by GPS's grouped sampling execution, making SDC the key link between classification results and sampling execution. The complete GARDEN configuration reflects the combined effect of cache, sampling, and batching co-design.}
\subsection{Scalability}
We evaluate GARDEN's scalability across multiple dimensions to demonstrate its robustness under various hardware configurations and workload characteristics. Unless specified otherwise, experiments use the setup described in Section 6.1: batch size 8000 on the PA dataset using a single GPU.
\textbf{GPU Scaling.} Figure~\ref{fig:scalability}a shows training throughput as we scale from 1 to 6 GPUs. For multi-GPU setups ($n \geq 2$), we include GNNLab, which allocates 1 GPU for sampling when $n \leq 4$ and 2 GPUs when $n > 4$. GARDEN consistently outperforms all baselines, with the performance gap widening as GPU count increases. This results from GARDEN's coordination of host-device communication and super-batch processing.

\textbf{Batch Size Scaling.} Figure~\ref{fig:scalability}b shows performance across batch sizes (4,000 to 12,000) on PA. GARDEN achieves 1.2× to 8.1× speedups over DGL, DUCATI, and FastGL, with larger batches yielding greater advantages. This trend stems from two factors: (1) increased node overlap across mini-batches enables more effective deduplication via our Super-batch Classifier, and (2) larger workloads allow the Grouping Parallel Sampler to achieve higher warp occupancy and more efficient memory coalescing.

\textbf{Feature Dimension Scaling.} As shown in Figure~\ref{fig:scalability}c, we vary feature dimensions (128 to 512) on PA, measuring both end-to-end training time (bars) and computation time only (lines). GARDEN maintains 1.2× to 11.3× speedups across all dimensions. The persistent gap between total and computation time confirms that our AHFCA effectively reduces data movement overhead even with larger features by strategically caching high-utility data.

\textbf{Graph Complexity Scaling.} Figure~\ref{fig:scalability}d evaluates increasingly complex sampling patterns ([5, 10], [5, 10, 15], [5, 5, 10, 10]) on PA. GARDEN achieves 1.2× to 8.18× speedups over baselines. This efficiency stems from (1) adjacency caching with offline reordering and (2) warp-level grouped sampling that converts random global accesses to coalesced shared memory operations. While FastGL matches GARDEN on simpler patterns like [5, 10], its performance degrades on complex patterns like [5, 5, 10, 10] where subgraphs exceed cache capacity, making sampling latency the bottleneck. In contrast, GARDEN's adaptive caching maintains low overhead regardless of subgraph size.

\subsection{Training Accuracy}
\label{subsec:Accuracy}

\begin{table}[!t]
  \caption{\zhulatest{Accuracy and epoch time of three-layer GraphSAGE node classification on PA and IGB.}}
  \label{tab:performance}
  \centering
  \footnotesize
  \setlength{\tabcolsep}{3.5pt}
  \renewcommand{\arraystretch}{1.0}
  {\color{darkgreen}
  \begin{tabular}{@{}llrr@{}}
    \toprule
    Dataset & System & \multicolumn{1}{c}{Accuracy} & \multicolumn{1}{c}{Epoch Time (s)} \\
    \midrule
    \multirow{4}{*}{PA} & DGL    & $0.670 \pm 0.003$ & $29.79$ \\
                        & DUCATI & $0.672 \pm 0.001$ & $4.86$ \\
                        & FastGL & $0.672 \pm 0.004$ & $15.32$ \\
                        & GARDEN & $0.671 \pm 0.003$ & $3.95$ \\
    \midrule
    \multirow{4}{*}{IGB} & DGL    & $0.581 \pm 0.002$ & $31.13$ \\
                         & DUCATI & $0.581 \pm 0.004$ & $7.99$ \\
                         & FastGL & $0.583 \pm 0.004$ & $19.17$ \\
                         & GARDEN & $0.582 \pm 0.001$ & $5.30$ \\
    \bottomrule
  \end{tabular}}
\end{table}

\zhulatest{We compare GraphSAGE training on PA and IGB using DGL, DUCATI, FastGL, and GARDEN. As shown in Table~\ref{tab:performance}, GARDEN achieves accuracy close to the baselines on both datasets: 0.671 $\pm$ 0.003 on PA and 0.582 $\pm$ 0.001 on IGB. Its epoch times are 3.95~s and 5.3~s, respectively, lower than those of the compared systems on the same datasets. These results indicate that GARDEN's cache, sampling, and batching optimizations accelerate training without changing the model training objective or sampling semantics. UK and UK-Union are excluded from this accuracy comparison because official node features are unavailable; their randomly generated features are used only for runtime and scalability evaluation.}

\section{Related Work}

\textbf{Sample-based GNN Training.} GraphSAGE \cite{GraphSAGE} introduced fixed-size neighbor sampling and a general aggregation framework for inductive learning on large graphs. Subsequent efforts target sampling bottlenecks: LazyGCN \cite{lazyGCN} reuses sampled subgraphs across epochs to lower sampling frequency while maintaining accuracy. GPU-accelerated samplers like NextDoor \cite{2021Nextdoor}, WholeGraph \cite{wholegraph}, and gSampler \cite{2023gSampler} optimize subgraph extraction through parallel algorithms. NeutronOrch \cite{2023NeutronOrch} and NeutronHeter \cite{NeutronHeter} employ CPU offloading and workload orchestration in heterogeneous systems to balance resources and enhance throughput.

\textbf{Cache-based GNN Training.} To mitigate feature gathering overheads, Quiver \cite{Quiver} implement GPU feature caching to minimize CPU-GPU transfers. GNNLab \cite{2022GNNLab} uses a unified feature cache for node loading but omits adjacency structures, limiting sampling efficiency. Embedding caches in SANCUS \cite{Sancus}, NeutronOrch \cite{2023NeutronOrch}, and FreshGNN \cite{freshgnn} reduce aggregation data movement. FastGL \cite{2024fastgl} combines memory-aware pipelines with sampling and loading optimizations for higher throughput. DUCATI \cite{2023DUCATI} extends caching to include both features and topology on GPUs, improving allocation under diverse workloads. For graphs exceeding memory capacity, DiskGNN \cite{DiskGNN} enables efficient out-of-core training by optimizing disk I/O to eliminate read amplification while maintaining full model accuracy.

\textbf{Micro-batch-based GNN Training.} To tackle the GPU memory wall for large-scale GNN training, Betty~\cite{Betty} proposes batch-level graph partitioning that decomposes mini-batches into redundancy-aware micro-batches, enabling deeper models and larger graphs under limited GPU memory with gradient accumulation. Building on this idea, Buffalo~\cite{Buffalo} further proposes memory-efficient node bucketing with flexible mappings and lightweight organization, reducing partitioning overhead, fragmentation, and redundant feature loading. Cherry~\cite{Cherry} addresses the high cost of data preparation by constructing a Global Message-passing Flow Graph, which eliminates repetitive per-epoch partitioning and improves load balance while reducing node redundancy. More recently, PDGNN~\cite{PDGNN} targets micro-batch training with degree-pruned partitioning and explicit redundancy elimination, pruning low-contribution edges and duplicated accesses to improve efficiency in small micro-batches and multi-GPU settings. Overall, these systems alleviate memory pressure by partitioning graphs to support larger batches or deeper models, but they primarily optimize computation partitioning within and across devices rather than irregular sampling latency, dynamic dual-cache allocation under strict single-GPU budgets, or PCIe-efficient sampling workflows.

GARDEN differs from existing approaches by co-optimizing caching, sampling, and batching in a hardware-aware framework. It achieves higher hit rates and lower latency through adaptive allocation, grouped parallelism, and super-batch classification, enabling efficient billion-scale training in settings where prior systems remain bottlenecked by I/O or compute imbalance.

\section{Conclusion}
In this paper, we tackle the critical challenge of scaling GNN training to billion-edge graphs, where the GPU memory wall imposes severe limitations on efficiency. Existing sampling-based frameworks struggle with excessive PCIe data transfers and irregular memory access patterns, leading to poor bandwidth utilization and computational inefficiency. To overcome these bottlenecks, we introduce GARDEN, a GPU-accelerated sampling-based GNN system that co-designs caching, sampling, and batching through three key innovations: the Adaptive High-Fidelity Cache Allocator for dynamic memory management under fixed budgets, the Grouping-based Parallel Sampler for optimized warp occupancy on power-law graphs, and the Super-batch Degree Classifier for cache-friendly data transfers. Extensive evaluations on billion-scale graphs demonstrate that GARDEN achieves significant speedups over the state-of-the-art baselines while maintaining model convergence and accuracy.

\bibliographystyle{IEEEtran}
\bibliography{sample}











\newpage

 




\vfill

\end{document}